# Emissions and Energy Impacts of the Inflation Reduction Act

**Authors:** John Bistline[1]*, Geoffrey Blanford[1], Maxwell Brown[2], Dallas Burtraw[3], Maya Domeshek[3], Jamil Farbes[4], Allen Fawcett[5], Anne Hamilton[2], Jesse Jenkins[6], Ryan Jones[4], Ben King[7], Hannah Kolus[7], John Larsen[7], Amanda Levin[8], Megan Mahajan[9], Cara Marcy[5], Erin Mayfield[10], James McFarland[5], Haewon McJeon[11], Robbie Orvis[9], Neha Patankar[12], Kevin Rennert[3], Christopher Roney[1], Nicholas Roy[3], Greg Schivley[13], Daniel Steinberg[2], Nadejda Victor[14], Shelley Wenzel[9], John Weyant[15], Ryan Wiser[16], Mei Yuan[17], Alicia Zhao[11]

[1] Electric Power Research Institute; Palo Alto, CA, USA.
[2] National Renewable Energy Laboratory; Golden, CO, USA.
[3] Resources for the Future; Washington DC, USA.
[4] Evolved Energy Research; San Francisco, CA, USA.
[5] U.S. Environmental Protection Agency; Washington DC, USA.
[6] Princeton University; Princeton, NJ, USA.
[7] Rhodium Group; Washington DC, USA.
[8] Natural Resources Defense Council; Washington DC, USA.
[9] Energy Innovation; San Francisco, CA, USA.
[10] Dartmouth College; Hanover, NH, USA.
[11] Center for Global Sustainability, University of Maryland; College Park, MD, USA.
[12] Binghamton University; Binghamton, NY, USA.
[13] Carbon Impact Consulting; New Providence, NJ, USA.
[14] National Energy Technology Laboratory; Pittsburgh, PA, USA.
[15] Stanford University; Stanford, CA, USA.
[16] Lawrence Berkeley National Laboratory; Berkeley, CA, USA.
[17] MIT Joint Program on the Science and Policy of Global Change; Cambridge, MA, USA.

*Corresponding author: John Bistline, jbistline@epri.com

Introduction

If goals set under the Paris Agreement are met, the world may hold warming well below 2 °C (*1*); however, parties are not on track to deliver these commitments (*2*), increasing focus on policy implementation to close the gap between ambition and action. Recently, the US government passed its most prominent piece of climate legislation to date—the Inflation Reduction Act of 2022 (IRA)—designed to invest in a wide range of programs that, among other provisions, incentivize clean energy and carbon management, encourage electrification and efficiency measures, reduce methane emissions, promote domestic supply chains, and address environmental justice concerns (*3*). IRA's scope and complexity make modeling important to understand impacts on emissions and energy systems. We leverage results from nine independent, state-of-the-art models to examine potential implications of key IRA provisions, showing economy wide emissions reductions between 43-48% below 2005 by 2035.

This multi-model analysis provides a range of decision-relevant information. For example, international policymakers and negotiators need to track progress toward Paris Agreement pledges, and assessing IRA's impacts is important to monitor US efforts and to provide a template for measuring the performance of other sectors and jurisdictions. Federal and state policymakers can use this IRA analysis to compare updated baselines with policy targets—for emissions, electric vehicle deployment, and others—to understand the magnitude of additional policies and private sector actions needed to narrow implementation gaps. Electric companies need to know how long IRA incentives will be available, since these subsidies can continue until electricity emissions are below 25% of their 2022 levels, which requires national models to evaluate. Industry- and technology-specific deployment can support investors, technology developers, researchers, and companies to quantify market opportunities.

Modeling IRA Provisions

Some of the models used in our analysis informed legislative debates preceding IRA's passage (*4-7*). We build on these preliminary analyses by updating IRA representations, increasing the number of models, and providing a systematic comparison of decision-relevant metrics. Multi-model studies highlight the robustness of insights and potential uncertainties to alternate model structures and input assumptions. Models in this study vary in their coverage and implementation of IRA provisions (Table S1). These differences are due to the models' scopes (e.g., 6 models of the full US energy system vs. 3 that focus on the power sector only) and resolutions (e.g., level of technological and sectoral detail). Variations in IRA implementation across models are also caused by the bill's complexity and pending guidance from government agencies, which require subjective judgments from modeling teams. The partial coverage of IRA provisions could imply that models underestimate emissions reductions, though many of IRA's largest provisions are represented and the degree of additionality of the remaining provisions is unclear (SM S7). Other simplifications (e.g., limited representations of frictions associated with infrastructure deployment, supply chains, and non-cost barriers) could result in higher emissions relative to modeled outcomes. These uncertainties imply that model results should not be interpreted as predictions (SM S1).

To evaluate impacts on emissions and energy systems, IRA scenarios are compared to their counterfactual reference scenarios without IRA (SM S2). IRA scenarios focus on central estimates of climate and energy provisions, which are not harmonized across models.

Emission Reduction Pathways

Economy-wide emissions reductions from IRA are 33-40% below 2005 in 2030 across multi-sector models with a 37% average (Fig. 1). This reflects a range of IRA provisions modeled, input assumptions, and model structures (SM S1). The 2030 range with IRA of 33-40% is a significant reduction from the reference without IRA incentives, which is 25-31% below 2005 (28% average). Emissions reductions from IRA grow over time and lead to 43-48% declines by 2035 from 2005 (compared with 27-35% in the reference). IRA helps to narrow the implementation gap in achieving the US 2030 target to reduce net greenhouse gas (GHG) emissions by at least 50% (*8-9*). The emissions gap is 1.0-1.6 Gt-$CO_2$e/yr without IRA, falling to 0.5 to 1.1 Gt-$CO_2$e/yr with IRA (19-25 percentage points to 10-17 p.p., Fig. S6).

Emissions reductions are not evenly distributed across sectors (Fig. 1B). Most IRA-induced mitigation comes from electricity, representing 38-80% of 2030 reductions (64% average) from the reference in economy-wide models. There is consistency across models that IRA will accelerate power sector decarbonization (Fig. S7). In 2030, power sector emissions with IRA are 47-83% (68% average) below their 2005 levels compared to 41-60% (51% average) in the reference. IRA-induced reductions continue through 2035, and the range narrows to 66-87% (77% average) with a 13-36 p.p. reduction from the reference (Fig. S7B), short of the goal of 100% "carbon pollution-free electricity" by 2035 *(8)*. The technology-neutral tax credits under IRA for zero-emitting resources continue after 2032 until electricity emissions are 25% of 2022 levels. Three of nine models reach this threshold by 2035.

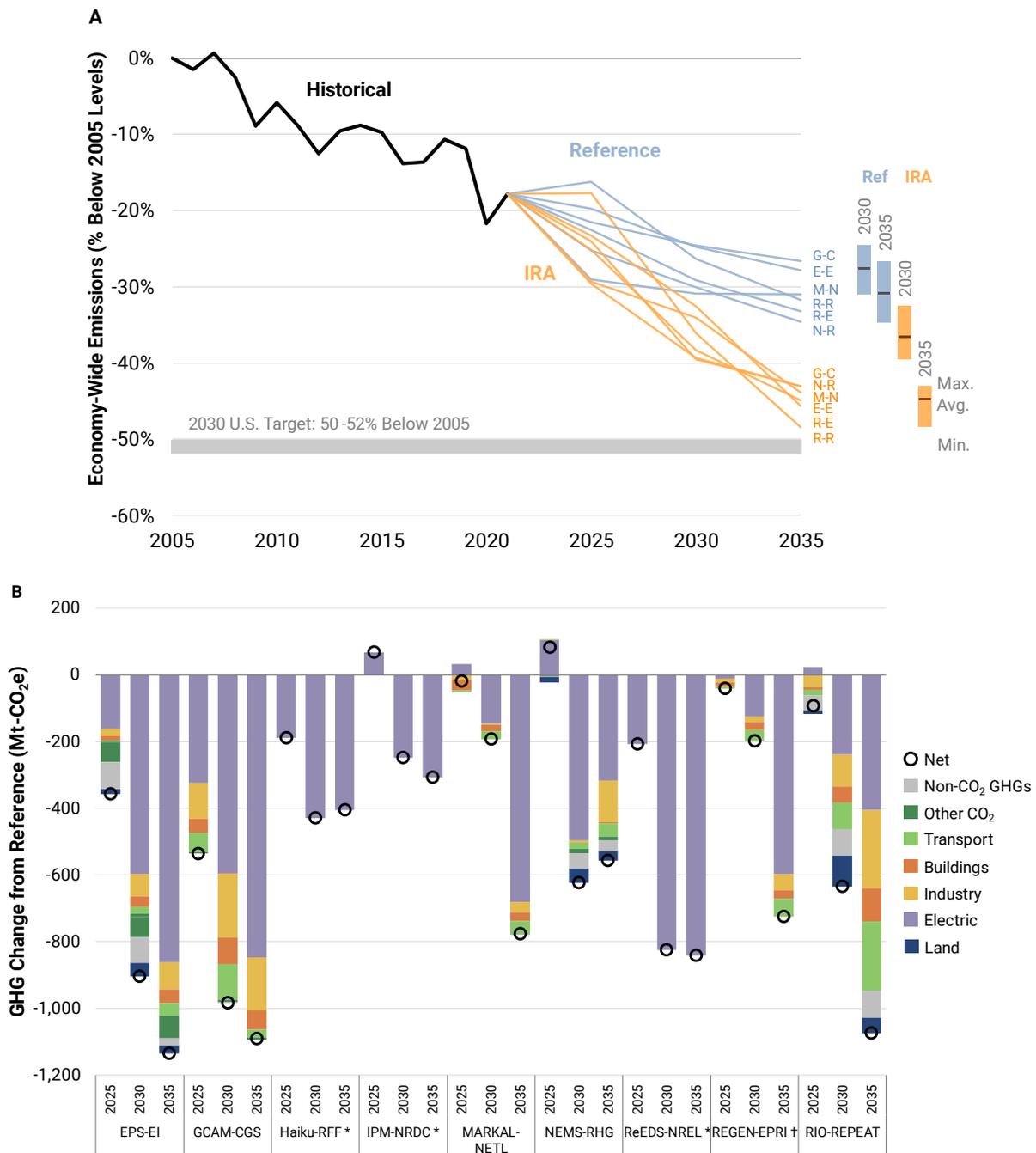

**Fig. 1. Cross-model comparison of U.S. emissions reductions under IRA and reference scenarios from 2005 levels.** **(A)** Historical and projected economy-wide GHG emissions. Historical emissions and 100-year Global Warming Potential values (for models representing non-CO$_2$ GHGs) are based on the U.S. EPA's "Inventory of U.S. Greenhouse Gas Emissions and Sinks." **(B)** Emissions reductions by sector and model over time under IRA scenarios relative to reference levels. Models with * designate that electric sector IRA provisions only are represented, and † denotes energy CO$_2$ IRA provisions only. Additional information on participating models and study assumptions can be found in Materials and Methods S1 and S2.

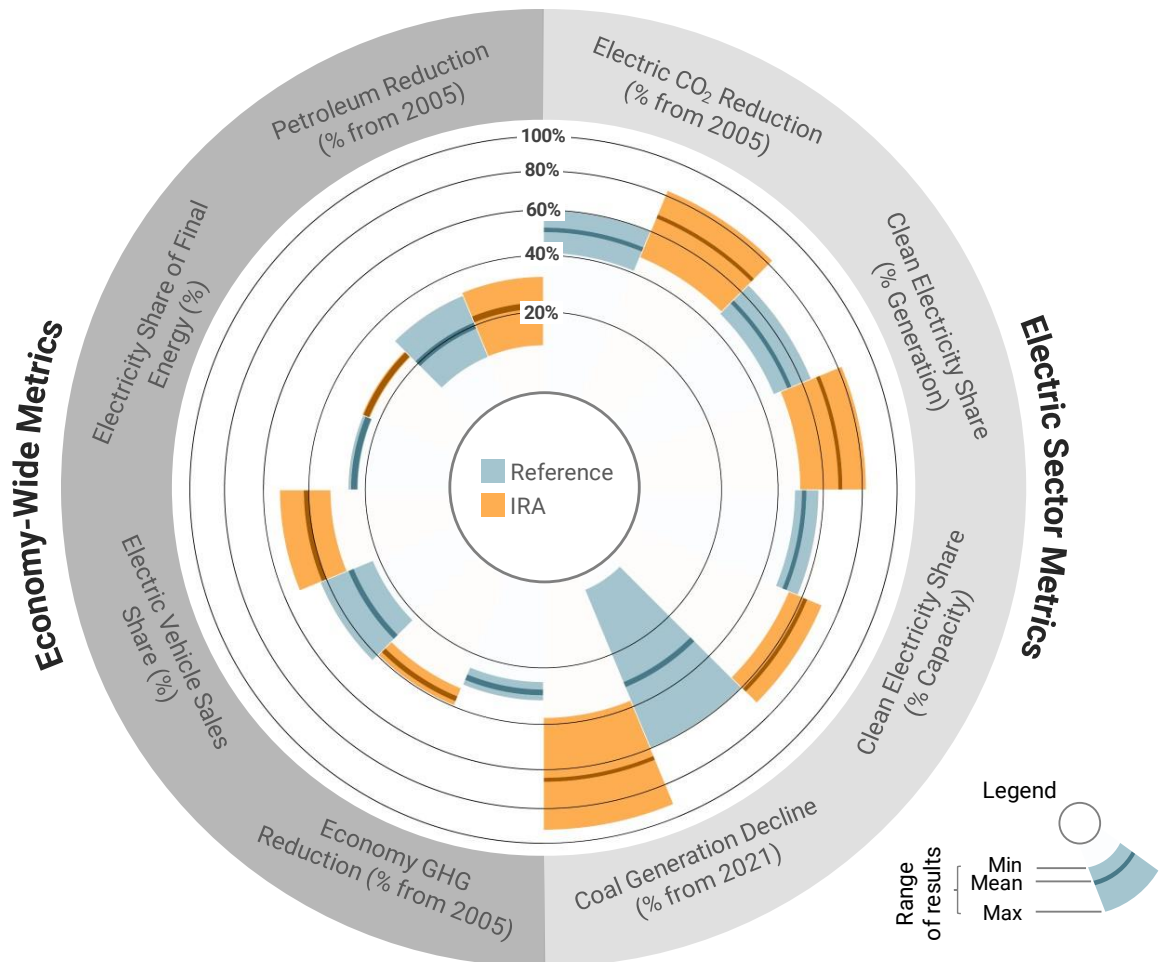

**Fig. 2. Summary of key indicators for IRA and reference scenarios across models.**
Clockwise from the top, indicators are 2030 electric sector $CO_2$ reductions (% from 2005 levels), 2030 generation share from low-emitting technologies (%, including renewables, nuclear, and CCS-equipped generation), 2030 capacity share from low-emitting technologies (% installed capacity), 2030 coal generation decline (% from 2021 levels), economy-wide $CO_2$ reduction (% from 2005 levels), 2030 electric vehicle new sales share (% of new vehicle sold are battery or plug-in hybrid electric), 2030 electricity share of final energy (%), and 2030 petroleum reduction (% from 2005 levels). Model-specific values for these metrics are provided in Table S6.

Effects of IRA on the Power Sector

IRA includes many electricity-related provisions (Table S1), including investment (30%) and production ($27.5/MWh, 10 year) tax credits for clean electricity resources, tax credits for energy storage and carbon capture, and tax credits to maintain existing nuclear plants. Some provisions involve long-term extensions of pre-IRA tax credits (e.g., for wind and solar), some involve increases in tax credit levels (e.g., carbon capture credits, bonuses for production and investment credits), and others are new (e.g., support for existing nuclear).

Models consistently show that IRA leads to large increases in wind and solar deployment but with substantial variation in magnitudes (Fig. S13A). Across all models, growth rates from 2021 to 2035 range from 10-99 GW/yr for solar and wind under IRA (58 GW/yr average), which is more than twice the average of 27 GW/yr without IRA and higher than the record 33 GW installed in 2021. There is wide variation in the expected increase in energy storage across models, 1-18 GW/yr (7 GW/yr average), compared with 0-8 GW/yr in the reference.

Results also exhibit reductions of unabated coal generation (i.e., without any carbon capture and storage (CCS)), ranging from 38-92% declines from 2021 levels by 2030 with IRA (Fig. 2) versus 3-60% without IRA. Five models show retrofits of some share of coal capacity with CCS, driven by the high value of tax credits for stored $CO_2$ (increasing from \$50/t-$CO_2$ historically to \$85/t-$CO_2$) (Fig. S13). Most models suggest that natural gas generation will decline under IRA relative to today's levels; however, gas-fired capacity increases in all models to provide firm capacity as coal retires and load grows (Fig. S13).

Overall, generation shares from low-emitting technologies—including renewables, nuclear, and CCS—in 2030 vary be-tween 49-82% (68% average) across models with IRA (Fig. S13B), up from about 40% to-day and from 46-65% without IRA (54% average), an 11-33 p.p. increase. Power sector generation and emissions outcomes under IRA are more closely aligned by 2035 across models (Fig. S7 and S13).

Implications of IRA on End-Use Demand

IRA reduces transport emissions by accelerating electrification. Across models, electric vehicles (EVs) are 32-52% of new light-duty vehicle sales by 2030 with IRA (41% average), compared with 22-43% (31% average) in the reference (Fig. 2 and S14), which is several times 2022 sales levels of about 7%. EV sales decrease or flatten between 2030 and 2035 due to the expiration of IRA tax credits. The electrified service demand share, stock share, and emissions lag new sales, given the turnover rate of the fleet (Fig. S15). EV credits in IRA also exist for commercial vehicles, including vans, buses, and trucks. These incentives help electrify these segments, representing about one-third of transport electricity demand by 2030 (Fig. S17). Transport exhibits the highest growth in electrification, though the magnitude varies by model (2-6% of final energy use in the sector by 2030 with IRA, 1-4% in the reference).

IRA incentives also encourage building efficiency and electrification, especially the adoption of heat pumps for space and water heating. Buildings currently have the highest electrification share of end-use sectors (just under 50% of final energy use in the sector), which increases as end-uses electrify (Fig. S16). Fuel switching from fossil fuels to electricity in buildings, transport, and industrial sectors leads electricity's economy-wide share of final energy to increase from about 21% today to 23-26% by 2030 (Fig. 2) and 25-29% by 2035.

These projected changes in demand may drive the first sustained period of declining petroleum use in US history (Fig. S18). By 2030, IRA scenarios show a 11-32% (22% average) decrease in petroleum consumption from 2005, but much of this happens in the reference (11-31% with 20% average) due to the competitiveness of transport electrification and continued efficiency improvement in the remaining internal combustion engine fleet. Natural gas consumption also decreases with IRA relative to the reference scenario, especially in the power sector, but declines relative to its historical peak are not as high as for petroleum and coal. IRA may catalyze several new markets in industry and fuels, including CCS, hydrogen, biofuels, and sustainable aviation fuels. Tax credits for captured $CO_2$ may accelerate emissions reductions across a variety of industries and lead to CCS deployment in the industrial sector, fuel production, and the power sector (Fig. S22).

Societal Implications of IRA

The climate-related benefits of IRA are estimated to be substantial (Fig. S10), even though there is uncertainty about the magnitude of the social cost of $CO_2$ (*10*, Fig. S9) and model-specific emissions impacts of IRA (Fig. 1B). Climate benefits range from $44-220B annually by 2030 across models using central social cost of $CO_2$ values with a 2% near-term discount rate ($20-100B with a 3% rate). Implied average abatement costs of IRA incentives per unit of $CO_2$ reduced range from $27-102/t-$CO_2$ with an average of $61/t-$CO_2$ across all models (Fig. S11). Although these costs are higher for economy-wide models (SM S4), abatement costs are much lower than many updated social cost of $CO_2$ estimates, even before accounting for co-benefits.

Declining fossil fuel use not only lowers GHG emissions but also conventional air pollutants, which improves public health outcomes. Fig. S12 compares reductions in $SO_2$ and $NO_x$ in the power sector and across the economy. Individual studies indicate that monetized health benefits from power sector air pollution reductions alone are $9-22B annually by 2030 (*11*, *12*), and $53B annually by 2030 from particulate matter reductions across the economy (*8*).

These declines in fossil fuel use generally mean that IRA lowers energy costs for households and businesses (Fig. S19), despite increases in electricity spending. The magnitudes of these consumer cost changes vary over time and across the four economy-wide models reporting fuel expenditures, but net spending declines $2-26B/yr by 2030 ($13-190 per household) relative to the reference and $10-52B per year by 2035 ($73-370 per household). Electrification and accelerated investments increase electricity expenditures for most economy-wide models, even though there is a reduction in total net energy service costs. Incentives for the adoption of end-use technologies such as EVs and heat pumps also can lower capital and maintenance costs.

The magnitude and composition of tax credit value under IRA vary by model (Fig. S21). Estimates from economy-wide models suggest that $330-870B could be spent on tax credits through 2030 ($510B average), which suggests uptake of IRA incentives could be greater than initial CBO/JCT estimates indicated (*3*). Cumulative tax credit expenditures through 2035 span $640-1,300B ($910B average), indicating that some IRA impacts may take longer to materialize. Individual studies perform analysis to understand broader societal implications of IRA, including improving distributional outcomes (*11*), increasing jobs (*5*, *6*, *12*), and bolstering domestic manufacturing (*13*).

Conclusions and Key Unknowns

While IRA accelerates decarbonization, including beyond 2030, no models indicate that the 2030 US climate target would be met with IRA alone. Overall, the analysis suggests that IRA may have its largest effects in the power sector, as its incentives amplify trends already underway and lower decarbonization costs. Modeling IRA impacts is challenging, since net effects depend on clean energy adoption, producer choices, household purchases, and actions by policymakers. Several key unknowns remain.

For example, lowering clean energy costs may increase ambitions of other federal agencies, state and local governments, and companies, though there are many complex drivers of such goals.

These dynamic effects of ratcheting ambition are not accounted for in the modeled scenarios described here but may be key to closing the 2030 implementation gap. In recent months, the US Environmental Protection Agency has released proposed standards that target emissions from cars, trucks, and power plants. These complementary regulations are aimed at leveraging IRA incentives to further accelerate decarbonization trends and lower costs of future regulations.

Also, expanded and new tax credits in IRA—combined with grants, loans, fees, domestic manufacturing incentives, and Infrastructure Bill incentives—support technologies that have experienced recent growth (e.g., wind, solar, batteries) and catalyze new markets for other decarbonization options (e.g., CCS, new nuclear, hydrogen, biofuels). Making clean technologies cheap to accelerate adoption can, in turn, buy down learning curves and encourage further deployment. By making low-emitting technologies cheaper domestically, IRA may have international spillovers, bolstering the political economy of those making their own investments and policies. Emerging technologies have larger uncertainty about induced technical change owing to their limited deployment and nascent markets.

Finally, models attempt to capture many economic factors that could influence technology adoption, but several implementation challenges are difficult to model, including the scale-up of supply chains and materials, siting and permitting, infrastructure expansion, network effects, non-cost barriers to consumer uptake of incentives, and the economic incidence of subsidies. These questions have prompted debate about permitting reform to streamline the infrastructure approval process and increase the pace of clean energy deployment, though legislative disagreements have emerged about the types of projects included and about balancing these reforms with environmental safeguards and community participation. Broader macroeconomic trends of escalating interest rates, materials costs, and labor costs can lower decarbonization rates, though IRA incentives can help to offset these factors (*14*). Additional analysis is important for understanding potential impacts of partial coverage of IRA provisions, IRA implementation uncertainties, as well as uncertainties about external factors, including inflationary trends, domestic macroeconomic environment, and global drivers (SM S2).

**Acknowledgments:** The views and opinions expressed in this paper are those of the authors alone and do not necessarily represent those of their respective institutions, the US Department of Energy (DOE), the US Government, or other funding agencies, and no official endorsement should be inferred. The authors thank Jordan Wingenroth for his assistance with social cost of $CO_2$ data. The AAAS recognizes the US Government's non-exclusive rights to use the Work for non-commercial, governmental purposes where such rights are established in the contract.

**Funding:** J.F., J.J., R.J., E.M., N.P., and G.S. were funded by the William and Flora Hewlett Foundation. B.K., H.K., and J.L. were funded by Bloomberg Philanthropies, the William and Flora Hewlett Foundation, and the Heising-Simons Foundation. H.M and A.Z. were funded by Bloomberg Philanthropies. R.W. was funded by Lawrence Berkeley National Laboratory under Contract No. DE-AC02-05CH11231 with the US DOE. M.B., A.H., and D.S. were funded by the DOE Office of Policy through the National Renewable Energy Laboratory, operated by Alliance for Sustainable Energy, LLC, for the DOE under Contract No. DE-AC36-08GO28308.

**Author contributions:**

Conceptualization: All authors

Methodology: All authors

Investigation: All authors

Visualization: J.B.

Writing – original draft: J.B., M.B., A.F., C.M., J.M.

Writing – review & editing: All authors

**Competing interests:** Jesse Jenkins is part owner of DeSolve, LLC, which provides techno-economic analysis and decision support for clean energy technology ventures and investors. A list of clients can be found at https://www.linkedin.com/in/jessedjenkins. He serves on the advisory boards of Eavor Technologies Inc., a closed-loop geothermal technology company, and Rondo Energy, a provider of high-temperature thermal energy storage and industrial decarbonization solutions and has an equity interest in both companies. He also provides policy advisory services to Clean Air Task Force, a non-profit environmental advocacy group, and serves as a technical advisor to MUUS Climate Partners and Energy Impact Partners, both investors in early-stage climate technology companies. Ryan Wiser is a senior scientist at Lawrence Berkeley National Laboratory, on partial detail under contract to the U.S. Department of Energy. The other authors declare no competing interests.

**Data and materials availability:** All data and materials associated with the analysis are available at https://doi.org/10.5281/zenodo.7879732.

# Supplementary Materials for

## Emissions and Energy Impacts of the Inflation Reduction Act

John Bistline[1]*, Geoffrey Blanford[1], Maxwell Brown[2], Dallas Burtraw[3], Maya Domeshek[3], Jamil Farbes[4], Allen Fawcett[5], Anne Hamilton[2], Jesse Jenkins[6], Ryan Jones[4], Ben King[7], Hannah Kolus[7], John Larsen[7], Amanda Levin[8], Megan Mahajan[9], Cara Marcy[5], Erin Mayfield[10], James McFarland[5], Haewon McJeon[11], Robbie Orvis[9], Neha Patankar[12], Kevin Rennert[3], Christopher Roney[1], Nicholas Roy[3], Greg Schivley[13], Daniel Steinberg[2], Nadejda Victor[14], Shelley Wenzel[9], John Weyant[15], Ryan Wiser[16], Mei Yuan[17], Alicia Zhao[11]

[1] Electric Power Research Institute; Palo Alto, CA, USA.
[2] National Renewable Energy Laboratory; Golden, CO, USA.
[3] Resources for the Future; Washington DC, USA.
[4] Evolved Energy Research; San Francisco, CA, USA.
[5] U.S. Environmental Protection Agency; Washington DC, USA.
[6] Princeton University; Princeton, NJ, USA.
[7] Rhodium Group; Washington DC, USA.
[8] Natural Resources Defense Council; Washington DC, USA.
[9] Energy Innovation; San Francisco, CA, USA.
[10] Dartmouth College; Hanover, NH, USA.
[11] Center for Global Sustainability, University of Maryland; College Park, MD, USA.
[12] Binghamton University; Binghamton, NY, USA.
[13] Carbon Impact Consulting; New Providence, NJ, USA.
[14] National Energy Technology Laboratory; Pittsburgh, PA, USA.
[15] Stanford University; Stanford, CA, USA.
[16] Lawrence Berkeley National Laboratory; Berkeley, CA, USA.
[17] MIT Joint Program on the Science and Policy of Global Change; Cambridge, MA, USA.

*Corresponding author: John Bistline, jbistline@epri.com

**Acknowledgments**
The views and opinions expressed in this paper are those of the authors alone and do not necessarily represent those of their respective institutions, the U.S. Department of Energy (DOE), the U.S. Government, or other funding agencies, and no official endorsement should be inferred. The authors thank Jordan Wingenroth for his assistance with social cost of $CO_2$ data.

**Funding**
J.F., J.J., R.J., E.M., N.P., and G.S. were funded by the William and Flora Hewlett Foundation. B.K., H.K., and J.L. were funded by Bloomberg Philanthropies, the William and Flora Hewlett Foundation, and the Heising-Simons Foundation. H.M and A.Z. were funded by Bloomberg Philanthropies. R.W. was funded by Lawrence Berkeley National Laboratory under Contract No. DE-AC02-05CH11231 with the US DOE. M.B., A.H., and D.S. were funded by the DOE Office of Policy through the National Renewable Energy Laboratory, operated by Alliance for Sustainable Energy, LLC, for the DOE under Contract No. DE-AC36-08GO28308.

**Competing Interests**
Jesse Jenkins is part owner of DeSolve, LLC, which provides techno-economic analysis and decision support for clean energy technology ventures and investors. A list of clients can be found at https://www.linkedin.com/in/jessedjenkins. He serves on the advisory boards of Eavor Technologies Inc., a closed-loop geothermal technology company, and Rondo Energy, a provider of high-temperature thermal energy storage and industrial decarbonization solutions and has an equity interest in both companies. He also provides policy advisory services to Clean Air Task Force, a non-profit environmental advocacy group, and serves as a technical advisor to MUUS Climate Partners and Energy Impact Partners, both investors in early-stage climate technology companies. Ryan Wiser is a senior scientist at Lawrence Berkeley National Laboratory, on partial detail under contract to the U.S. Department of Energy. The other authors declare no competing interests.

**This PDF file includes:**

> Materials and Methods S1 to S7
> Figs. S1 to S26
> Tables S1 to S6
> References

**Materials and Methods**

S1: Participating Energy-Economic Models

The nine-model intercomparison includes a wide range of independent, state-of-the-art models. These models capture complicated economic and energy system interactions of the Inflation Reduction Act (IRA) and other policies. Model intercomparisons are used in a variety of fields to identify robust insights and potential areas of uncertainty (1).

Models in the study vary in their coverage and structure. Three models are partial equilibrium models that focus on the power sector, while the other six models represent broader energy systems. Power sector models can provide additional temporal, spatial, and technological detail for system operations and investments, while energy system models capture linkages with broader systems and the economy, including cross-sector interactions that are amplified by IRA's end-use electrification incentives. Table S2 and Table S3 compare key model features and provide links to detailed documentation. Table S4 and Table S5 compare model representations of emerging technologies and expansion constraints.

The nine participating models are:

- **EPS-EI:** The [Energy Policy Simulator](#) (EPS) is a forward-simulating, annual timestep, single-region model, developed by Energy Innovation LLC, which aims to provide information about which climate and energy policies will reduce GHG emissions most effectively and at the lowest cost. It includes every major sector of the economy: transportation, electricity supply, buildings, industry, agriculture, and land use. The EPS takes outputs from other publicly available models and studies, such as the U.S. Energy Information Administration's *Annual Energy Outlook*, to create a business-as-usual scenario. When users select policies, the model tracks changes from the business-as-usual projections to estimate how policy affects energy demand and costs, among other outputs.

- **GCAM-CGS:** GCAM-USA-AP is a special purpose fork of the [Global Change Analysis Model](#) (GCAM) version 5.3, utilized and maintained by the Center for Global Sustainability. GCAM-USA-AP follows the standard release of GCAM 5.3, but also adds detailed representations of sector-specific climate policies at the state level (2). GCAM tracks emissions of 16 different species of GHGs and air pollutants from energy, agriculture, land use, and other industrial systems. The energy system formulation in GCAM consists of detailed representations of depletable primary sources such as coal, gas, oil, and uranium, in addition to renewable resources such as bioenergy, hydropower, wind, and geothermal. These energy resources are processed and consumed by end users in the buildings, transportation, and industrial sectors. GCAM is a hierarchical market equilibrium model. The equilibrium in each period is solved by finding a set of market prices such that supplies and demands are equal in all simulated markets (3).

- **Haiku-RFF:** Resources for the Future's [Haiku](#) model is a system operation and capacity planning model of the U.S. electricity sector. With perfect foresight across a 26-year time horizon, it finds the least-cost way to meet electricity sector demand in each of the transmission-constrained 48 contiguous states and D.C., with representation of state-level market characteristics and technology mandates and emissions caps. System operation within each year is broken into three seasons with eight time blocks each. Fuel costs, load shapes, declining technology costs, and rising demand are exogenous.

- **IPM-NRDC:** The [Integrated Planning Model](#) (IPM) is a multi-regional, dynamic, and deterministic linear programming model of the U.S. electric power sector. IPM optimizes for the least-cost pathway available for the construction, economic retirement, and use of power plants, subject to resource adequacy requirements and environmental constraints. It is used by the U.S. Environmental Protection Agency (EPA) for regulatory impact assessments of power sector regulations, as well as by state agencies and the Regional Greenhouse Gas Initiative. It is a proprietary model of ICF.

- **MARKAL-NETL:** [MARKAL](#) is a bottom-up, dynamic, linear programming optimization model that finds the cost-optimal pathway within the context of the entire energy system. MARKAL does not contain an in-built database, so in this study, the publicly available EPAUS9r2017 database for the U.S. energy system has been adopted and modified by the National Energy Technology Laboratory (NETL). MARKAL-NETL represents U.S. Census regions from 2010-2075 with five-year time periods. Each of the nine regions is modeled as an independent energy system with different regional costs, resource availability, existing capacity, and end-use demands. Regions are connected through a trade network that allows transmission of electricity and transport of gas and fuels. Electricity transmission is constrained to reflect existing regional connections between North American Electric Reliability Corporation regions as closely as possible. MARKAL-NETL represents energy imports and exports, domestic production of fuels, fuel processing, infrastructures, secondary energy carriers, end-use technologies, and energy service demands of the entire economy.

- **NEMS-RHG:** RHG-NEMS is a version of the Energy Information Administration's [National Energy Modeling System](#) (NEMS) modified by Rhodium Group. RHG-NEMS is comprised of 13 modules providing energy sector-wide coverage on the supply and demand side as well as macroeconomic interactions and interactions with global energy markets. The supply-side modules generally rely on least-cost optimization, while the demand-side modules are a combination of least-cost optimization and other consumer adoption modeling approaches. Outside of NEMS, which provides energy $CO_2$ projections, Rhodium Group applies an in-house model to project the additional GHGs targeted for reduction under the Kyoto Protocol. Regionality, temporal resolution, and technology representation vary across modules.

- **ReEDS-NREL:** The U.S. Department of Energy National Renewable Energy Laboratory's [Regional Energy Deployment System](#) (ReEDS) is a publicly available, bottom-up representation of the U.S. electricity sector. In the setup for this study, the linear program portrays electricity supply and demand as well as the provision of operating reserves for grid reliability at 134 different balancing areas while also representing 356 sub-regions where variable renewable capacity can be built. The Augur sub-module solves hourly dispatch across multiple load years to estimate capacity credit and curtailment.

- **REGEN-EPRI:** The U.S. [Regional Economy, Greenhouse Gas, and Energy](#) (REGEN) model is developed and maintained by the Electric Power Research Institute (EPRI). REGEN links a detailed power sector planning and dispatch model with an energy end-use model (4). The power sector model simultaneously finds cost-minimizing pathways for capacity investments, transmission expansion, and dispatch. The model features hourly resolution to capture the evolving end-use mix and a representative hour approach for power sector investment and operations. The end-use model captures technology choices at the customer level with differences across model regions, sectors, and structural classes.

- **RIO-REPEAT:** The [Regional Investment and Operations Model](#) (RIO) supply-side model and EnergyPATHWAYS demand-side model were developed by Evolved Energy Research. The models provide detailed energy accounting and examine optimal energy system investment and operations. The tools have high resolution across sectors of the economy (more than 60 U.S. energy system subsectors), time (annual turnover of equipment stocks coupled with an hourly electricity-dispatch model), and geography (16 different regions of the United States along with the transmission connecting them). The modeling tools are employed to conduct a rigorous technical quantification of the infrastructure upgrades and technology investments needed across all sectors of the energy system, including cross-sectoral opportunities, to achieve both near- and long-term climate goals while meeting projected demand for energy services.

Results in the text compare outputs across these nine models in terms of ranges and mean values, which offer rough indications of variation across models as well as measures of central tendency. However, it is important to bear in mind that comparisons across models, much like scenario ensembles, should not be interpreted as statistical samples or as indications of the likelihoods of particular outcomes (5). Model results are not distributions and only provide an ad hoc representation of uncertainty, as the full uncertainty is likely to be larger than the range suggests (e.g., due to structural uncertainties associated with models as well as parametric uncertainties associated with future technologies, policies, and markets).

Model outputs should not be interpreted as predictions of policy-induced changes for several reasons. First, models vary in their coverage and implementation of IRA provisions (Table S1), as described in SM S2. The bill's complexity and pending guidance from government agencies require subjective judgments from modeling teams. Second, there is considerable uncertainty about technological change, policies, inflationary trends, domestic macroeconomic environment, and global drivers. Policy and technological change, which IRA may amplify, are particularly uncertain. Third, models vary in their scope and resolution, which may impact energy system and emissions outcomes. For instance, the literature indicates that model choices related to temporal and spatial resolution are linked to renewables and energy storage deployment (6; 7). While power sector impacts are explored with both economy-wide and power-sector-only models alike, impacts outside of the power sector are not evaluated with detailed sectoral models to understand how increased resolution may alter modeled outcomes.

S2: Scenario Design

IRA scenarios represent central estimates of core climate and energy provisions of each respective model. Models vary in their coverage and implementation of IRA provisions (Table S1). The core modeled provisions are expected to capture the majority of IRA's impacts, even though many provisions are not explicitly represented. Economy-wide and power sector models all include core extensions and enhancements of power-sector-related tax credits, including production tax credits (PTC), investment tax credits (ITC), and credits for captured $CO_2$ (45Q). Note that Table S1 is not an exhaustive list of IRA provisions, some of which are modeled while others are not. Eligibility requirements for tax credit bonuses aim to encourage high-road jobs, increase deployment of low-emitting technologies in low-income/energy, and spur domestic manufacturing. Many models assume that labor bonuses are met and that some technology-specific share of projects qualify for other bonuses.

To understand effects on emissions and energy systems, IRA scenarios are compared to their counterfactual reference scenarios without IRA but including other current federal and state policies, regulations, and incentives. Many models incorporate on-the-books policies in this benchmark scenario up through early 2022 such as:

- The Infrastructure Investment and Jobs Act
- Federal ITC, PTC, and 45Q with phase outs
- State-level clean electricity standards and renewable portfolio standards, including technology-specific carveouts and mandates (e.g., energy storage, offshore wind)
- Regional- and state-level emissions policies, including economy-wide policies (e.g., California's cap-and-trade) and power sector $CO_2$ caps (e.g., the Regional Greenhouse Gas Initiative)
- Federal and state performance standards for end-use technologies

These reference scenarios generally do not include proposed but not yet final regulations, including EPA's proposed methane rule and others.

This model intercomparison does not harmonize other technology, market, and policy assumptions, which means that models use their native input assumptions for technological costs.[1] Input assumptions for capital costs of key power sector resources are shown over time in Fig. S1.[2] Many models assume exogenous technological change, including reductions in capital costs of generation and energy storage options (Fig. S1).[3] Models broadly align in their trends of these technologies, and values over time generally fall in the range of the National Renewable Energy Laboratory's *Annual Technology Baseline*, which several use for their input assumptions. Table S4 compares model representations of emerging technologies, including carbon capture and storage (CCS), hydrogen, and carbon removal.[4] Table S5 provides model-specific assumptions about deployment constraints across different technologies. Many models include constraints on the near-term deployment on technologies with longer lead times (e.g., transmission, nuclear) over the next few years.

There is greater variation in natural gas prices across models (Fig. S2). This range reflects uncertainty about how prices of fossil fuels will change over time, especially given near-term inflationary drivers and the Russo-Ukrainian war.

S3: IRA Background

IRA is challenging to model due to its scope and complexity. The policy design of IRA through tax incentives, grants, loans, and rebates does not necessarily guarantee a fixed amount of emissions reductions or a price on emissions that encourages lowest marginal abatement cost opportunities. Unlike standards or emissions caps, IRA's investment-based climate policy approach does not target specific outcomes but instead provides an extension and expansion of previous investment-centered approaches.[5] In other words, IRA changes the relative prices of fuels and end-use equipment by making lower-emitting resources lower cost, but does not directly price carbon or cap emissions. Variation in IRA implementation across models is also caused by the bill's complexity and elements that require guidance from government agencies.

---

[1] Harmonization can isolate policy effects but does not capture the coevolution and impacts of policy, market, and technology drivers that could affect IRA performance.
[2] Monetary values are expressed in 2020 U.S. dollars unless otherwise noted.
[3] EPS-EI and NEMS-RHG include endogenous technological learning.
[4] All models generally include electricity generation options such as solar, wind, and natural-gas-fired capacity.

Updates to power sector tax credits include extending their timeline and value (e.g., increasing 45Q credits for captured and stored $CO_2$ from \$50/t-$CO_2$ to \$85/t-$CO_2$), expanding their eligibility (e.g., making the ITC and PTC technology-neutral, allowing standalone energy storage to claim the ITC) and flexibility (e.g., allowing technologies to claim the ITC or PTC, depending on which is more lucrative in their specific circumstances), adding bonus credits (e.g., for energy communities and domestic content), and improving access (e.g., direct pay for nonprofits and tax-exempt utilities and making credits transferrable for others). Tax equity market assumptions in models appear as reductions in effective ITC and PTC values. Future tax credit values are generally assumed to be indexed for inflation.

These complexities make modeling IRA challenging relative to earlier decarbonization policies and proposals. There is also considerable uncertainty in terms of how IRA could unfold and a range of possible outcomes based on different interpretations of its provisions (SM S7). Additionally, the dependence of several decarbonization pathways on infrastructure expansion—electricity transmission, hydrogen infrastructure, $CO_2$ pipelines, and others—raises questions about permitting and the degree to which such infrastructure may limit or accelerate technology adoption. Table S4 compares model representations of emerging technologies and their associated infrastructure, while Table S5 summarizes model-specific expansion constraints. Note that the IRA scenarios in the main text focuses on central estimates of climate and energy provisions. However, uncertainty about IRA implementation, combined with uncertainties about external factors, implies that ranges for outcomes of interest may be broader than the values in the main text.

S4: Emissions Results

Fig. S5B shows the sectoral emissions reductions under IRA scenarios relative to 2005, with Fig. S3 showing 2030 emissions levels in this scenario.[6] For non-$CO_2$ GHGs, note that Carbon Dioxide Equivalence ($CO_2e$) calculations use 100-year Global Warming Potential (GWP) values from IPCC's Fourth Assessment Report (AR4) to be consistent with the UNFCCC reporting requirements. Waterfall diagrams for individual models in Fig. S4 show emissions reductions from IRA relative to 2005 levels and the resulting emissions gap to reach 50% reductions by 2030. This emissions gap is 1.0-1.6 Gt-$CO_2e$/yr in the reference scenario and reduces to 0.5 to 1.1 Gt-$CO_2e$/yr with IRA.

Sectoral emissions reductions relative to the reference from IRA are shown in Fig. 1B. In 2025, most models indicate a reduction in emissions from IRA relative to the reference. However, a third of models exhibit an increase in power sector emissions in 2025 with IRA, since models are no longer front-loading wind and solar investments to take advantage of expiring tax credits. By 2030 (2035), IRA decreases emissions by 190-990 Mt-$CO_2$e (560-1,140 Mt-$CO_2$e) annually below the reference for the economy-wide models. Fig. 1 in the main text compares differences in economy-wide emissions between the IRA and reference scenarios over time, indicating that differences in 2030 emissions reductions are not solely due to differences in reference levels in 2030.

---

[5] Revenue from IRA comes from adjusting the minimum tax rate for corporations, taxing stock buybacks, enforcing existing taxes, and negotiating Medicare drug prices. Participating models do not represent potential feedbacks from these tax reforms.
[6] Emissions changes are relative to self-reported 2005 baselines to better harmonize reporting categories.

Power sector emissions over time are shown in Fig. S7 under the reference and IRA scenarios. The bottom panel of this figure illustrates the percentage point difference between the IRA and reference scenarios. IRA lower power sector emissions by 5-34 p.p. in 2030 and 13-36 p.p. in 2035. The scatter plot in Fig. S8 compares model-specific reductions in the reference and IRA scenarios and suggests that model responsiveness to IRA incentives is not driven primarily by different reference scenarios. Wind and solar deployment, fossil fuel declines, and other trends under the baseline play some role in IRA projections, but other factors including model structure (e.g., temporal resolution,[7] financing, foresight in Table S3) and input assumptions (e.g., technological costs in Fig. S1, natural gas prices in Fig. S2[8]) are also influential in cross-model variation in emissions and clean energy deployment. This figure also illustrates how IRA outcomes narrow through 2035 relative to 2030, largely due to models with lower abatement by 2030 nearing the other models by 2035.

One study estimates that net 2030 emissions increases from the oil and gas leasing provisions in IRA are unlikely to exceed 50 Mt-$CO_2$/yr (8). Even with this conservative assumption, for every ton of emissions increase from oil and gas provisions, there would be as much as 18 tons of emissions abated through other IRA provisions.

Climate benefits calculations in Fig. S10 are based on social cost of $CO_2$ distributions from the GIVE model in Rennert, et al. (2022) (9). Distributions for the social cost of $CO_2$ by discount rate are shown in Fig. S9 and are based on Resources for the Future Socioeconomic Projections (RFF-SP) scenario samples with uncertainty in climate model, sea-level model, and climate damage parameters. These values are multiplied by the difference in 2030 emissions with and without IRA (Fig. S5) to estimate the distribution of climate benefits, which represents reduced societal damages associated with agriculture, mortality, sea-level rise, energy consumption, and other impacts. Note that Fig. S10 indicates which models include only the electric sector so that their generally lower climate benefits can be attributed to their limited scope (rather than to their more limited emissions reductions). The full distribution of damage estimates in Fig. S10 imply a broad range of climate benefits.

Fig. S11 compares distributions of social cost of $CO_2$ estimates in 2030 with average abatement costs across models. These calculations take net changes in energy system costs in 2030 relative to the reference without IRA—including energy costs (Fig. S19), tax credits (Fig. S21), capital costs of supply- and demand-side technologies, and maintenance—and divide these costs by emissions reductions from IRA in 2030 (i.e., the difference between GHG emissions in the IRA case and the reference without IRA, Fig. S5). Incentives can deliver benefits far into the future by lowering operating costs of the energy system, so incorporating these longer time horizons to measure the savings would decrease abatement costs. Note that cost-effectiveness calculations use undiscounted costs for comparability with Joint Committee on Taxation (JCT) and Congressional Budget Office (CBO) values.[9] There are several other caveats to bear in mind in comparing average abatement costs with social cost of $CO_2$ estimates, including the omission of other co-benefits (e.g., improved air quality), the possible divergence between average and marginal costs,[10] and the exclusion of deadweight losses from market distortions.

---

[7] Temporal resolution refers to the number of intra-annual periods represented for electric sector investment and dispatch decisions, which are compared across models in Table S2. Earlier research indicates that the number of intra-annual timeslices and their selection can materially alter model assessments of the economics of power sector decarbonization (20).

[8] Several models assume lower natural gas price trajectories than recent levels. If higher prices persisted, regional natural gas shares and decarbonization would shift, potentially increasing reference decarbonization rates and reducing incremental impacts of IRA tax credits (21; 22).

Average abatement costs across models range from $27-102/t-$CO_2$ with an average of $61/t-$CO_2$ all models. Economy models have higher abatement costs ($71/t-$CO_2$ average) than electric sector models ($57/t-$CO_2$ average), given the cost-effectiveness of power sector tax credits vis-à-vis non-electric incentives under IRA. In part, abatement costs are higher for some end-use credits owing to the fraction of inframarginal transfers to households that would have adopted these IRA-supported technologies even without tax credits (e.g., electric vehicles in Fig. S14). Even with these transfers and higher fiscal costs (Fig. S21, discussed in S5), average abatement costs of IRA are generally lower than updated social cost of $CO_2$ values across many assumed near-term discount rates (Fig. S11). This comparison suggests that additional emissions reductions could be warranted, because mitigation costs are generally below ranges for the social cost of $CO_2$, even before accounting for improved air quality and other co-benefits.

As GHG emissions decline due to IRA incentives, other forms of air pollution from fossil fuel combustion also decline. Fig. S12 shows declines in economy-wide and power sector $NO_x$ and $SO_2$ emissions over time.[11] These emissions declines continue historical trends, though IRA scenarios have accelerated reductions relative to the reference without IRA.

S5: End-Use Results

Electricity demand grows 16% on average for energy system models with endogenous load growth between 2021 and 2030 under IRA (Fig. S17), which is higher than 13% on average in the reference case. Transportation leads demand growth, though electrification in buildings and industry also contribute (Fig. S16). The fraction of light-duty vehicle sales coming from EVs (including both battery electric and plug-in hybrid electric vehicles) increases from over 7% in 2022 to 32-52% of new sales by 2030 with IRA (41% average), compared with 22-43% (31% average) in the reference (Fig. S14). The $7,500 tax credit may prove restrictive in terms of its constraints based on domestic content, assembly, as well as price- and income-based eligibility. The electrified service demand share, stock share, and emissions lag new sales, given the turnover rate of the existing fleet (Fig. S15). Other transportation represents a growing share of transport electrification over time (Fig. S17) due in part to the fewer restrictions on credits for business and commercial vehicles.

IRA leads to declining fossil fuel consumption across most models and fuels relative to the reference scenario (Fig. S18), though magnitudes vary by model and fuel. The extent of petroleum reductions depends largely on substitution with electricity in the transport sector. Coal and natural gas consumption with IRA relative to the reference depend more on power sector investment and generation outcomes. Coal consumption continues to decline across most models, though a near-term rebound in coal use occurs for models with higher natural gas price assumptions and greater CCS deployment.[12]

---

[9] These comparisons also could compare discounted costs (i.e., to reflect the opportunity cost of capital) as well as discounted emissions (i.e., to reflect discounting of future climate-related benefits).
[10] Unlike carbon pricing or other policy instruments, estimating marginal abatement costs with tax credits requires more than simply reporting the shadow price on the emissions cap constraint (18; 19).
[11] Note that we do not use these emissions trajectories as inputs to air quality modeling and then assess changes in monetized damages across these scenarios. Estimating air quality co-benefits from IRA is an important area for future work.

IRA tends to lower energy costs for households and businesses due to declines in fossil fuel consumption and power sector subsidies (Fig. S19). Total economy-wide cost changes vary by model and over time. Economy-wide models indicate declines in petroleum and natural gas spending, which can be partially offset by increases in electricity expenditures. There is cross-model variation in whether IRA increases or decreases total power sector expenditures relative to a counterfactual reference without IRA. Increases could be due to greater electricity demand (from fuel switching), while decreases could be from lower electricity prices (due to investments in subsidized resources). The balance of these effects depends on which tax credits are used (e.g., investment credits lower upfront costs, while production credits lower operating costs over the first decade after an asset comes online), price formation in models, as well as capacity deployment and timing.

Fig. S20 shows how IRA decreases residential retail electricity prices in 2030 between 1-8% relative to the reference. Since electricity prices decline across all models, this result indicates that increased electricity expenditures in Fig. S19 are likely due to increased electricity demand from electrification. This hypothesis is supported by the fact that the three models indicating substantial IRA-induced reductions in power sector costs are partial equilibrium models, which do not account for quantity changes from electrification.

IRA tax credit values are shown in Fig. S21. Initial estimates of IRA funding and federal budgetary effects by the JCT and CBO between Fiscal Years 2022 and 2031 indicated nearly $400B for climate- and energy-related programs, including about $270B for power sector, fuels, and end-use credits (10). The utilization of IRA tax credits across models in this analysis suggests a broader range of potential tax expenditures, ranging from $330-870B in total by 2030 ($510B average) across economy-wide models, which is 1.2-3.2 times the CBO/JCT score for comparable credits. Large shares of IRA spending are allocated to provisions with uncapped incentives, including production- and investment-based tax credits. Tax credit values are higher across many categories and models in this analysis relative to the CBO/JCT estimates, though transport-related credits[13] and 45Q credits for captured $CO_2$ are categories with low CBO/JCT estimates and potentially large contributions in several models (Fig. S21). The analysis finds larger budget impacts after 2031 with cumulative spending of $640-1,300B ($910B average) by 2035 due to extensions of power sector tax credits. Additional public spending on advanced manufacturing and other climate-related incentives are not included in this total. If these economy-wide tax expenditures were combined with direct expenditures[14] in IRA ($121B), total

---

[12] In particular, MARKAL-NETL exhibits an increase in coal consumption due to its high CCS deployment in the power sector with 45Q credits, which entails parasitic energy penalties relative to coal capacity without $CO_2$ capture.

[13] Uncertainty about the magnitude of clean vehicle credits is due not only to unknowns about future electric vehicle sales but also to uncertainty about the share of vehicles qualifying for critical minerals and battery sourcing requirements, eligibility restrictions, and whether leased vehicles are exempt from these stringent requirements. Note that other studies such as Cole, et al. (forthcoming) (17) estimate clean vehicle credits to have budgetary effects of about $450B through 2031, which is more than an order of magnitude larger than the CBO/JCT score.

[14] Direct expenditures include agricultural and forestry projects, energy loans, industrial decarbonization funds, Green Bank, and several others.

fiscal costs over the ten-year budget window would be $450-1,000B ($630B average). There is a mix across credit types by model, though production tax credits are the largest expenditure category for many models. Per Fig. S11 and SM S4, average mitigation costs of IRA, including these expenditures, are generally below ranges for the social cost of $CO_2$. The range of values across economy-wide and electric sector models reflects both the range of possible technology adoptions as well as uncertainty about credit eligibility and magnitudes of bonus credits.

IRA could strengthen these areas where the U.S. already has comparative advantages in geologic storage, fossil fuel resources, and technical expertise. Fig. S22 shows how tax credits for captured $CO_2$ (45Q) may lead to CCS deployment in the industrial sector, fuel production, carbon removal, and the power sector. The extent of captured and stored $CO_2$ varies considerably by model and over time. Total annual volumes of captured $CO_2$ range from 10-350 Mt-$CO_2$/yr in 2030 (150 Mt-$CO_2$/yr average) and 10-810 Mt-$CO_2$/yr in 2035 (280 Mt-$CO_2$/yr average). Fig. S23 illustrates how tax credits for hydrogen (45V), alongside credits for captured $CO_2$, increase low-emissions hydrogen production. Hydrogen production shares vary by model, but IRA tax credits generally increase electrolytic hydrogen and CCS-equipped production.[15]

IRA manufacturing credits may bolster domestic production of solar modules, wind turbines, and batteries. Initial estimates suggest that these manufacturing subsidies could displace demand with domestically sourced products and even switch the U.S. to become a net exporter in these areas, though other countries may respond with tariffs or WTO challenges (11). Many of the models in this analysis do not capture these manufacturing credits or represent international market dynamics (Table S1).

S6: Land Use and Non-$CO_2$ GHG Emissions

Enhancing the U.S. land sink—also referred to as land use, land use change, and forestry (LULUCF)—and lowering non-$CO_2$ GHG emissions are additional mitigation pathways from IRA, which can contribute toward the 2030 climate target (12). Fig. S24 shows the net negative emissions from the U.S. land sink in 2030 across models. The sink ranges from -750 to -850 Mt-$CO_2$e (-800 Mt-$CO_2$e average). The land sink is larger than reference projections for each model and is comparable to the 2030 land sink range in the "United States 7th UNFCCC National Communication, 3rd and 4th Biennial Report" (13), which indicates a land sink between -720 and -860 Mt-$CO_2$e absent additional policies.

IRA also contributes to non-$CO_2$ GHG mitigation. Fig. S24 shows how these emissions decline under IRA relative to their reference levels. Most of the economy-wide models represent the Methane Emissions Reduction Program in IRA (Table S1), which includes a methane emissions fee and financial support to monitor and reduce methane associated with oil and natural gas systems. Upstream emissions associated with oil and gas also decline due to decreasing activity in these sectors under IRA (Fig. S18).

S7: Sensitivity Analysis

Given how models vary in their coverage of IRA provisions (Table S1), we conduct a sensitivity to illustrate how the core provisions that all models capture are the ones driving the largest energy system and emissions changes. This "IRA Core" scenario turns off the provisions that not all models capture and includes production, investment, existing nuclear, and captured $CO_2$ tax credits.[16]

---

[15] Impacts of 45V credits on production and emissions depend in part on Treasury guidance, which was forthcoming at the time this analysis was conducted.

Fig. S25 compares emissions, capacity additions, and capacity retirements across these scenarios. Power sector emissions decline 79-87% by 2035 from 2005 levels with scenarios with full IRA provisions and 75-86% with core IRA provisions only, indicating that the non-core provisions only lower power emissions by 1-4 percentage points across models. Likewise, electric sector capacity additions and retirements are similar across the full and core IRA scenarios, suggesting that changes are driven primarily by the core provisions that all participating groups model.

We also conduct low and high IRA sensitivities to understand how assumptions about implementation can alter emissions. These scenarios use alternate assumptions about bonus credit eligibility, end-use tax credit eligibility, and implementation details while holding all other assumptions constant (e.g., input assumptions about technological cost and performance).

Emissions reductions across these scenarios are shown in Fig. S26. The range of emissions outcomes is broader if IRA implementation uncertainties are included. For economy-wide models, GHG reductions by 2035 with IRA are 43-48% below 2005 levels in the middle IRA scenarios presented earlier, and this range increases to 38-51% when the low and high sensitivities are included. The power sector is where the largest changes occur in these sensitivities—electricity-related emissions reductions with IRA are 66-87% below 2005 by 2035 in the middle IRA scenarios (across all models), which increases to 60-92% across the low, middle, and high IRA implementation sensitivities. The responsiveness to changes in IRA implementation varies across models and is asymmetric in some instances between the low and high sensitivities. Ultimately, the greater inter-model variation across the middle IRA scenarios relative to intra-model variation across IRA sensitivities underscores the value of model intercomparison studies in understanding robust insights and possible variation.

---

[16] Specifically, the "IRA Core" scenario omits provisions related to solar in low-income communities, accelerated depreciation, funds for rural coops, and transmission financing. Some models exclude these provisions, since their model structures and scopes do not allow these incentives to be explicitly represented. This sensitivity focuses on the power sector, given how all models capture these provisions and how the electricity sector is the primary area for IRA-induced emissions reductions.

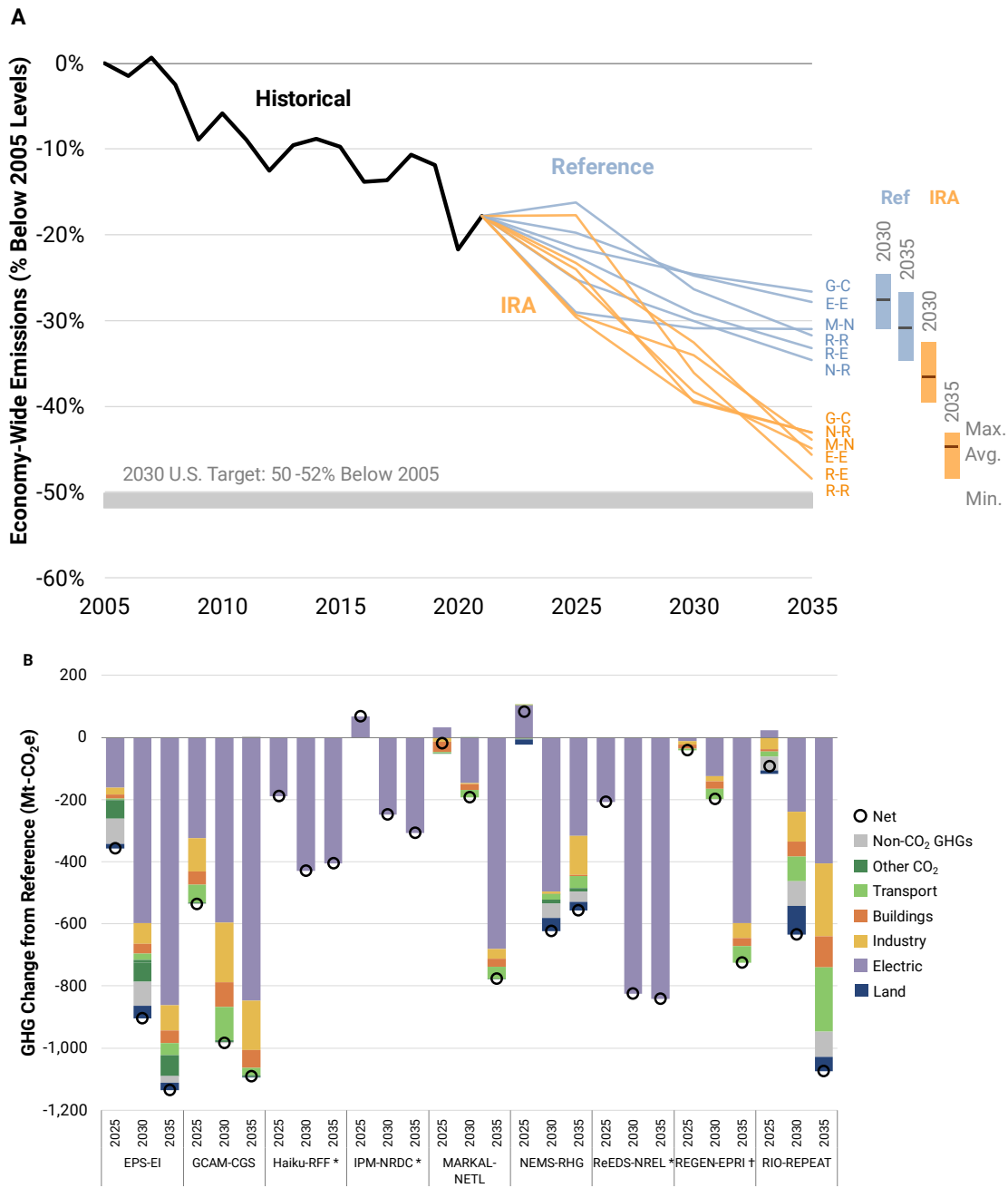

**Fig. 1. Cross-model comparison of U.S. emissions reductions under IRA and reference scenarios from 2005 levels.** (**A**) Historical and projected economy-wide GHG emissions. Historical emissions and 100-year Global Warming Potential values (for models representing non-$CO_2$ GHGs) are based on the U.S. EPA's "Inventory of U.S. Greenhouse Gas Emissions and Sinks." (**B**) Emissions reductions by sector and model over time under IRA scenarios relative to reference levels. Models with * designate that electric sector IRA provisions only are represented, and † denotes energy $CO_2$ IRA provisions only. Additional information on participating models and study assumptions can be found in Materials and Methods S1 and S2.

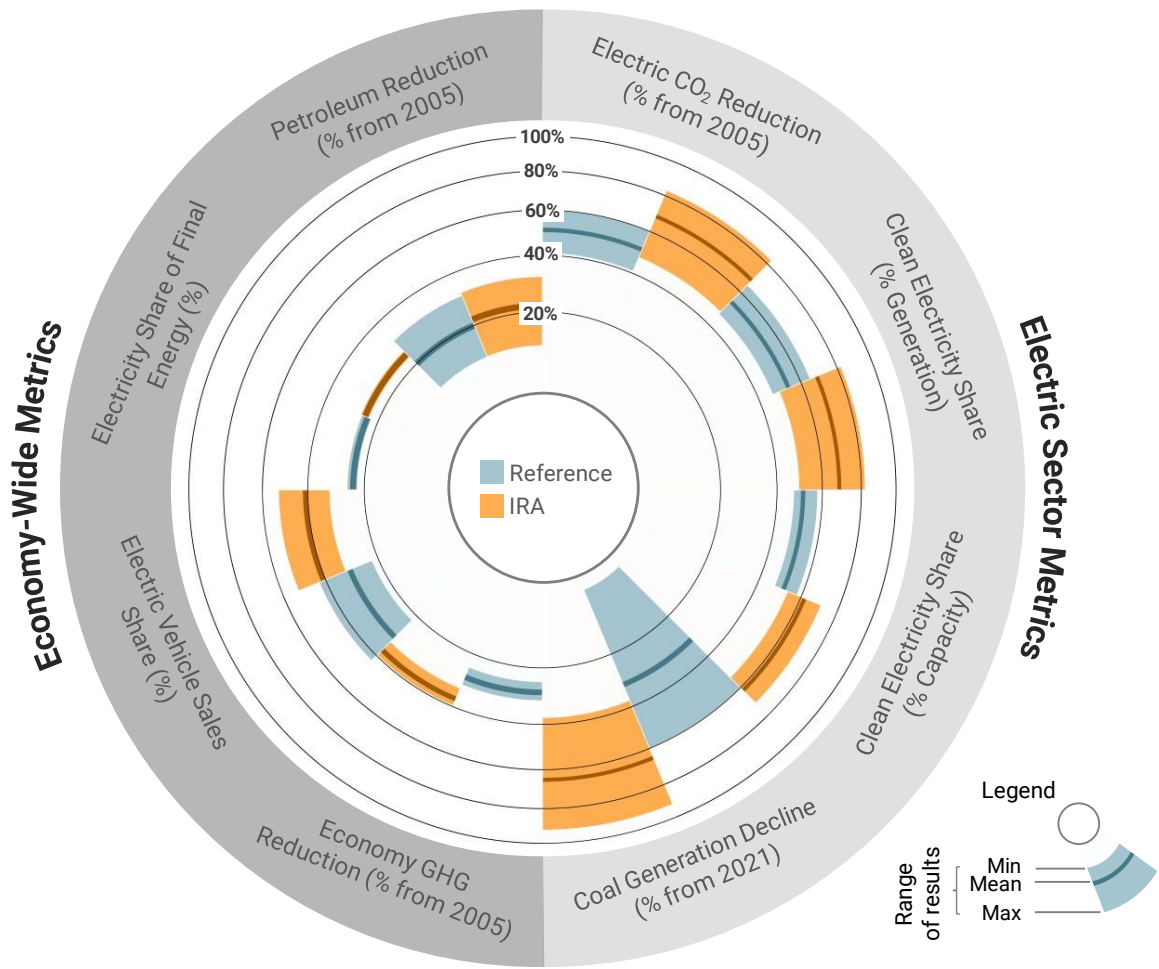

**Fig. 2. Summary of key indicators for IRA and reference scenarios across models.**
Indicators are 2030 electric sector $CO_2$ reductions (% from 2005 levels), 2030 generation share from low-emitting technologies (%, including renewables, nuclear, and CCS-equipped generation), 2030 capacity share from low-emitting technologies (% installed nameplate capacity), 2030 coal generation decline (% from 2021 levels), economy-wide $CO_2$ reduction (% from 2005 levels), 2030 electric vehicle new sales share (% of new vehicle sold are battery or plug-in hybrid electric), 2030 electricity share of final energy (%), and 2030 petroleum reduction (% from 2005 levels). Model-specific values for these metrics are provided in Table S6.

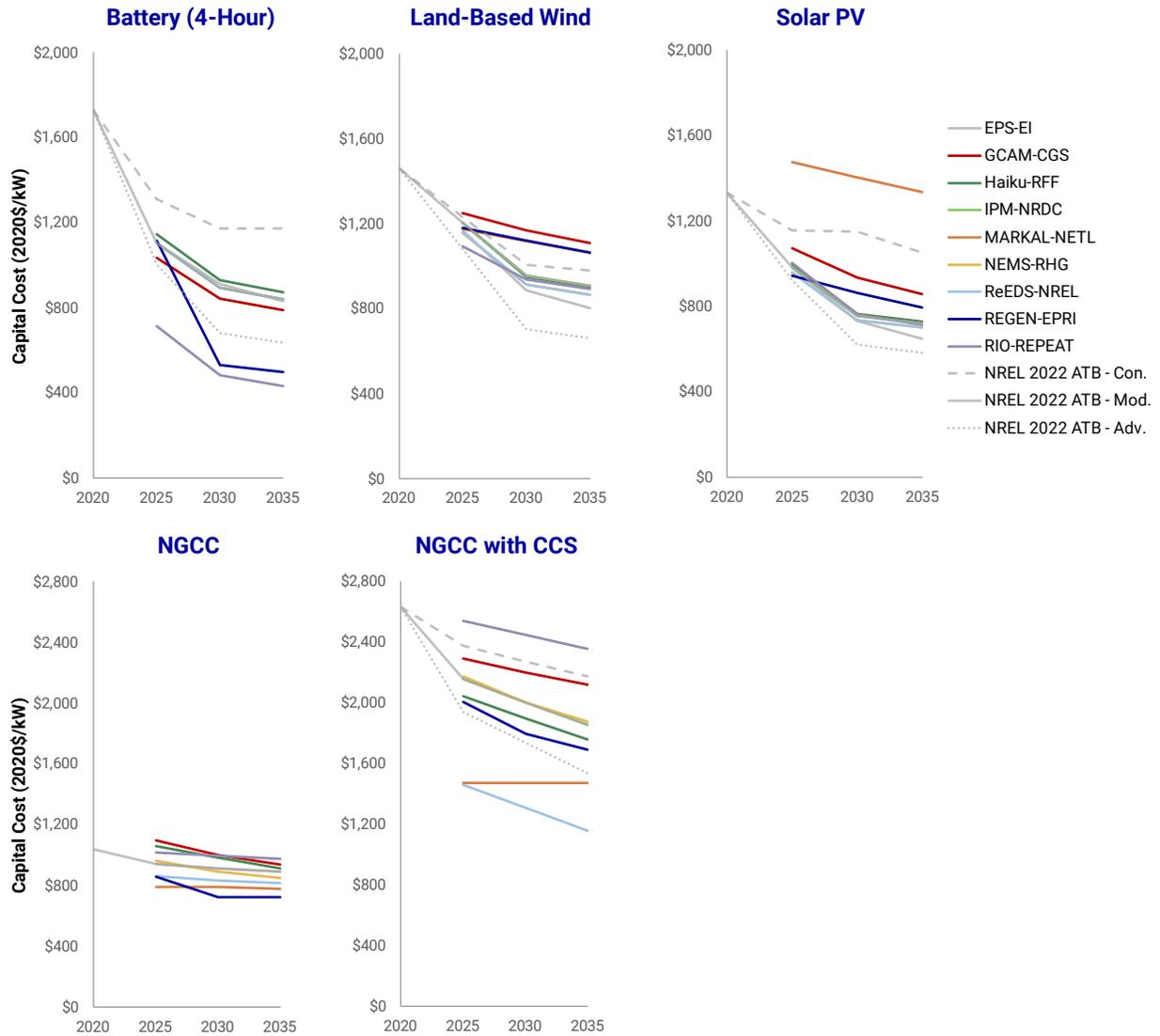

**Fig. S1. Capital cost assumptions of key power sector technologies over time by model.**
Costs are shown in 2020 U.S. dollars per kilowatt of nameplate capacity (utility-scale solar PV capacity is shown in $kW_{AC}$ terms).

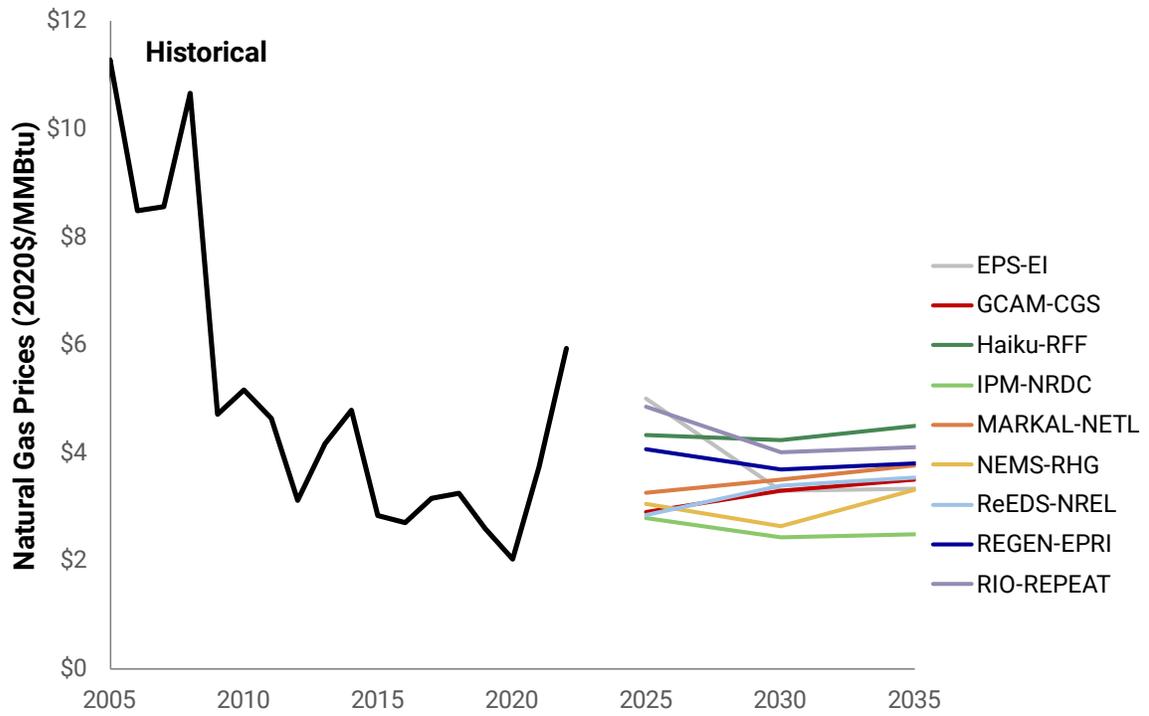

**Fig. S2. Natural gas price assumptions over time by model.** Henry Hub prices are shown where available (or delivered prices to the power sector).

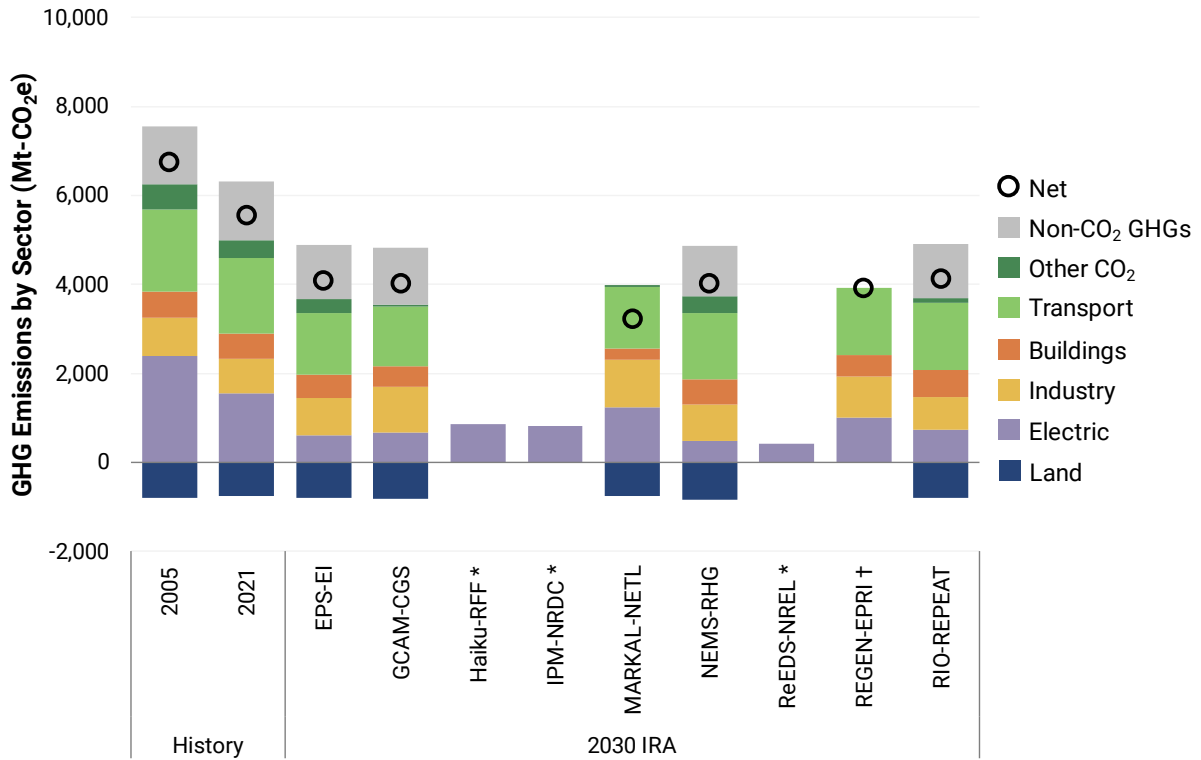

**Fig. S3. Cross-model comparison of U.S. GHG emissions by sector under IRA.** Historical emissions and 100-year Global Warming Potential values are based on the U.S. EPA's "Inventory of U.S. Greenhouse Gas Emissions and Sinks." Electric, industry, buildings, and transport show $CO_2$ only. "Non-$CO_2$ GHGs" includes other GHGs across all sectors. Models with * designate that electric sector IRA provisions only are represented ($CO_2$ only), and † denotes energy $CO_2$ IRA provisions only.

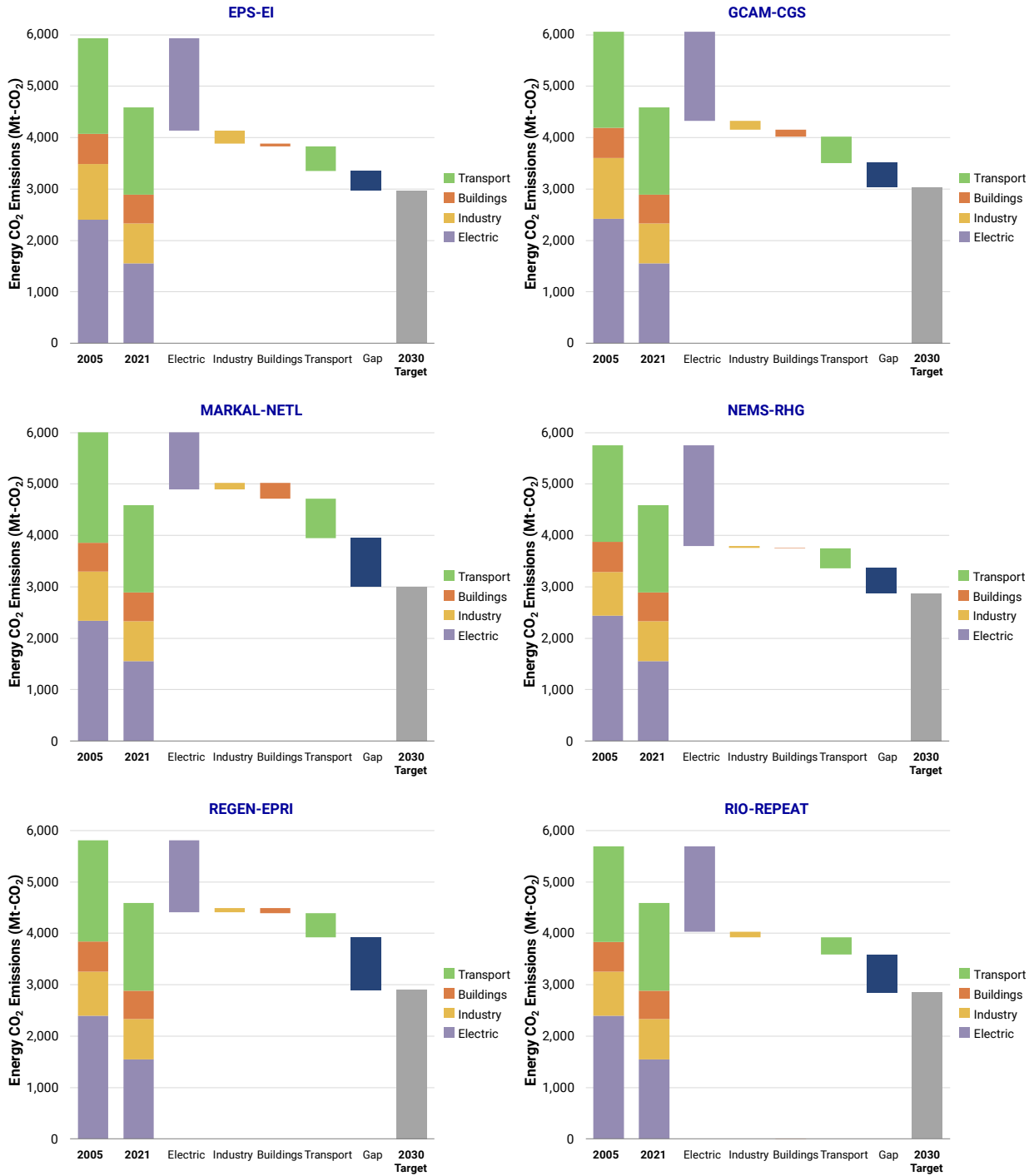

**Fig. S4. Cross-model waterfall diagrams of energy CO$_2$ emissions under IRA relative to 2005.** Panels show model-specific emissions under IRA and the emissions gap (navy bar) to reach a 50% by 2030 climate target. 2005 values are based on self-reported baseline values.

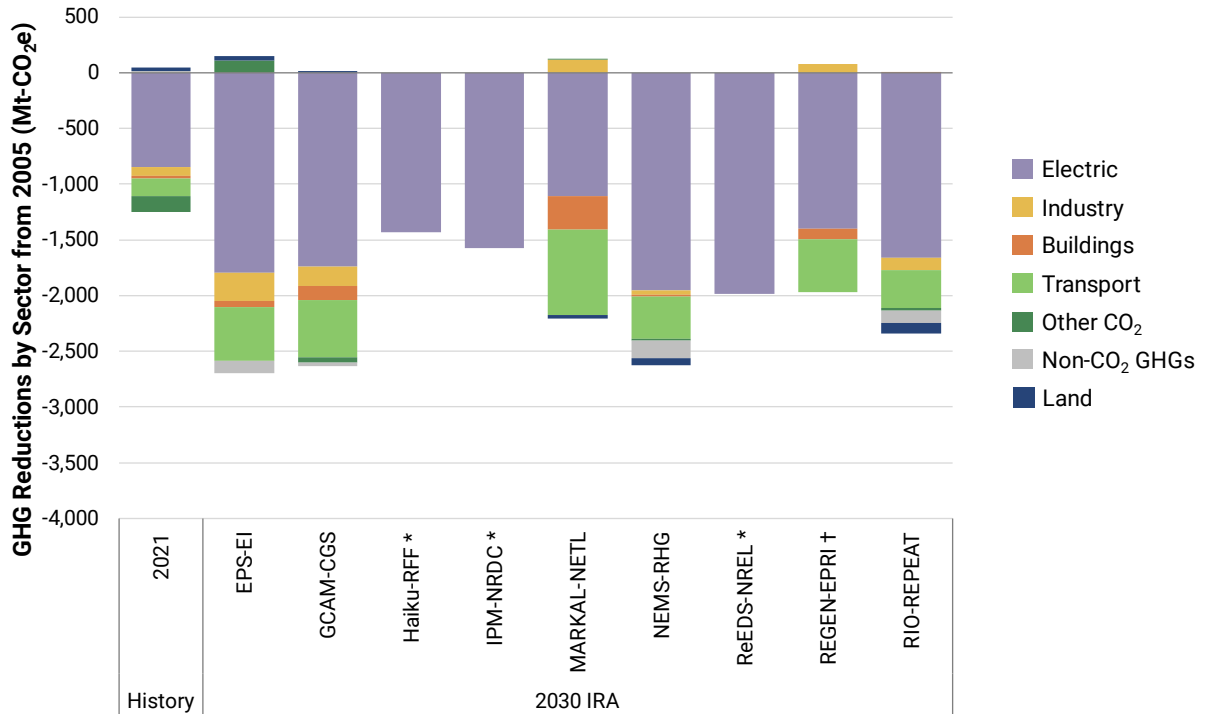

**Fig. S5. Emissions reductions by sector and model in 2030 under IRA scenarios from 2005 levels.** Models with * designate that electric sector IRA provisions only are represented, and † denotes energy $CO_2$ IRA provisions only.

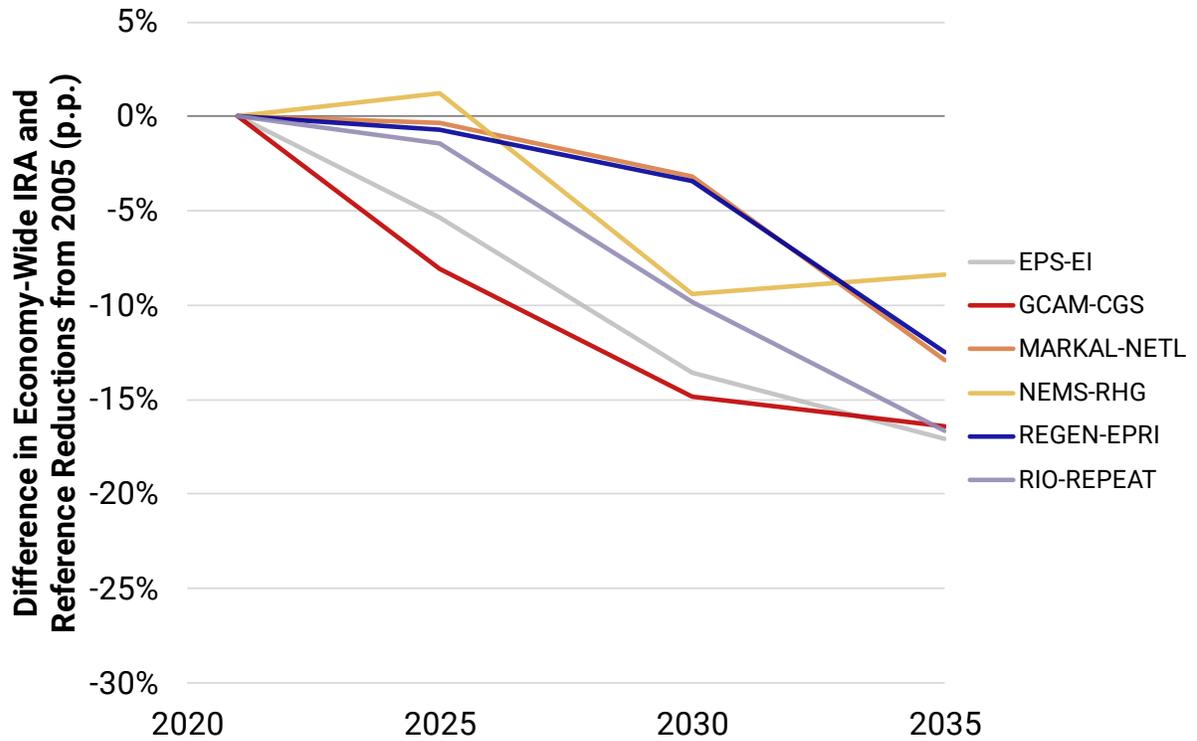

**Fig. S6. Difference in economy-wide GHG emissions reductions between the IRA and reference scenarios from 2005 levels (in percentage point terms).** Differences corresponding to trajectories over time in Fig. 1A. Note that economy-wide GHG emissions reductions relative to 2005 are calculated for each model using only the emissions sources (i.e., energy $CO_2$, land $CO_2$, other $CO_2$, and non-$CO_2$ GHGs) reported for that model, as shown in Fig. S3.

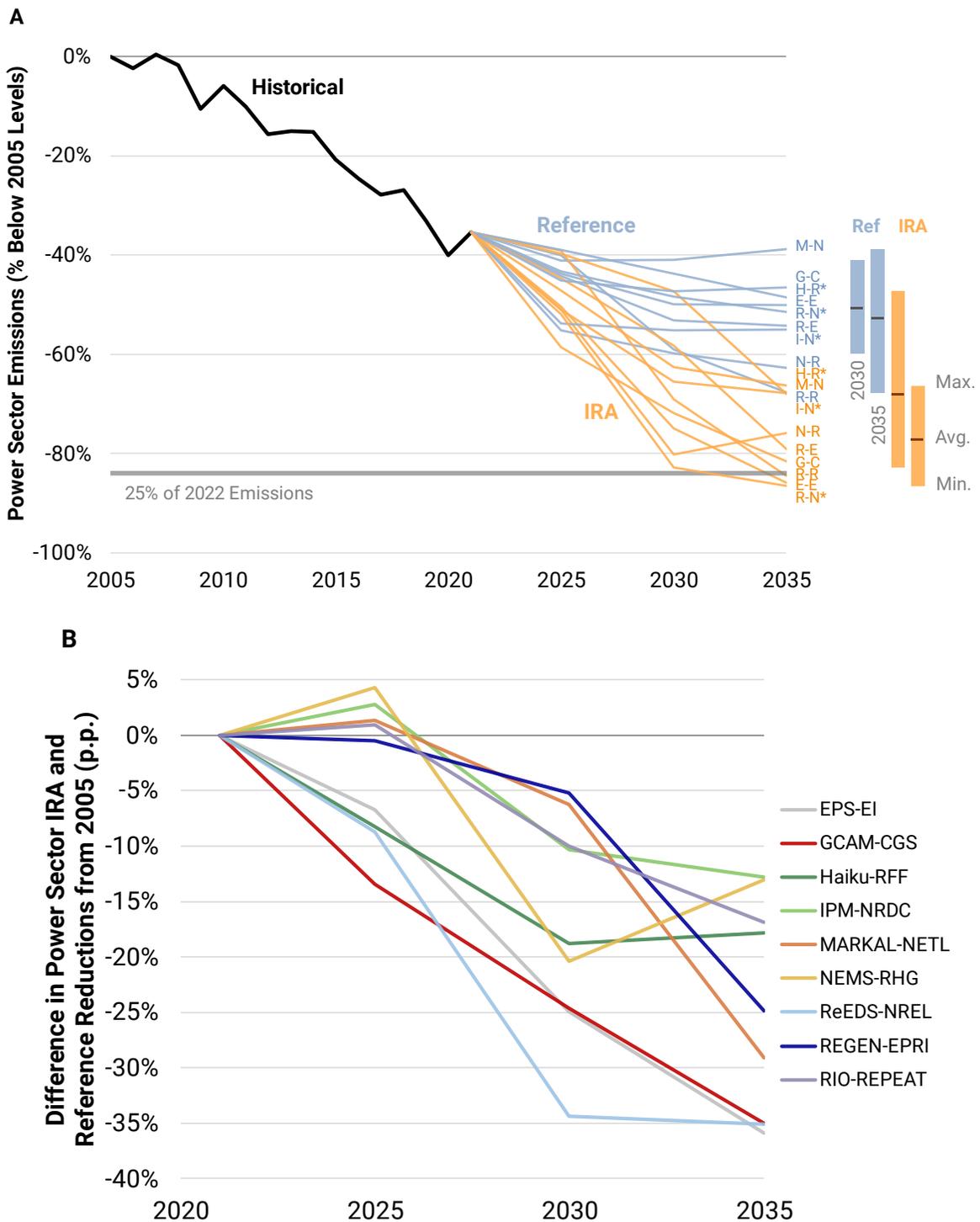

**Fig. S7. Cross-model comparison of U.S. power sector emissions reductions under IRA and reference scenarios.** (**A**) Reductions below 2005 levels. (**B**) Difference between the IRA and reference scenarios (in percentage point terms). Historical emissions are based on the U.S. EPA's "Inventory of U.S. Greenhouse Gas Emissions and Sinks." Models with * designate that electric sector IRA provisions only are represented.

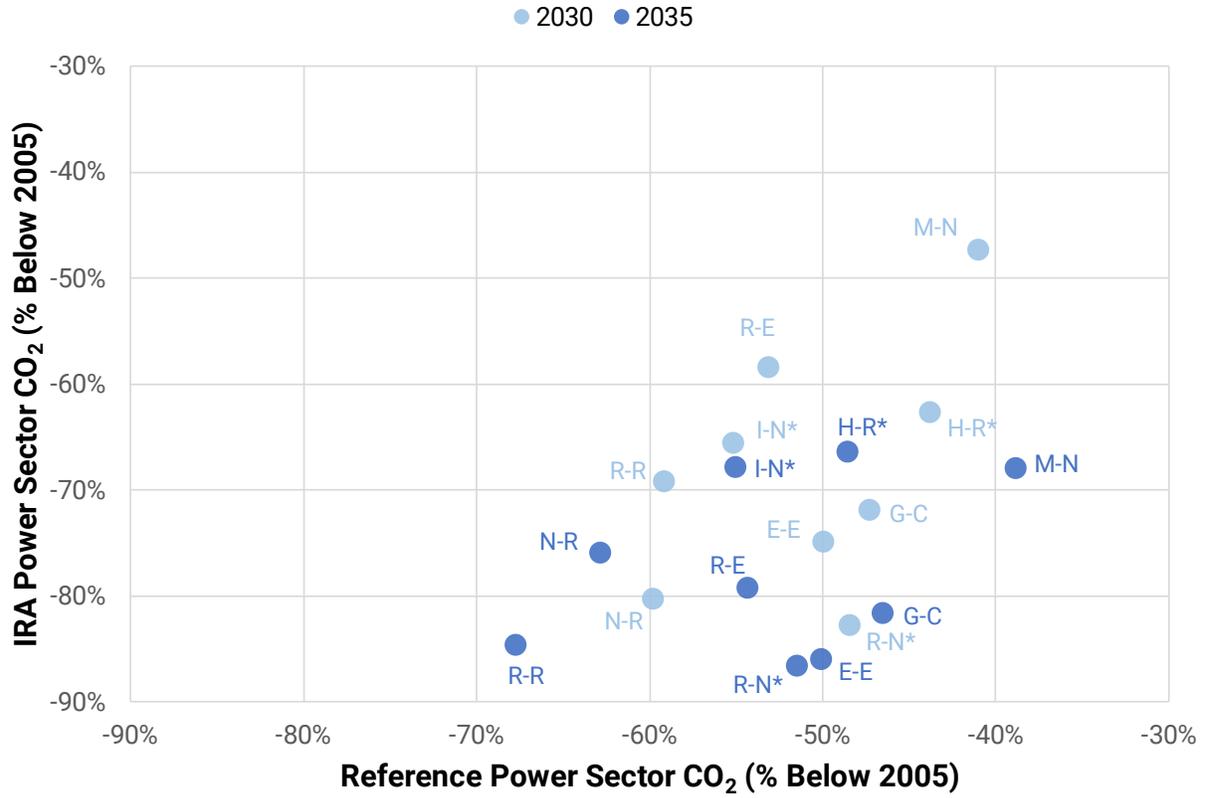

**Fig. S8. Cross-model comparison of power sector CO₂ emissions reductions (relative to 2005 levels) under the reference and IRA scenarios.** Individual model results are shown for 2030 and 2035. Models with * designate that electric sector IRA provisions only are represented.

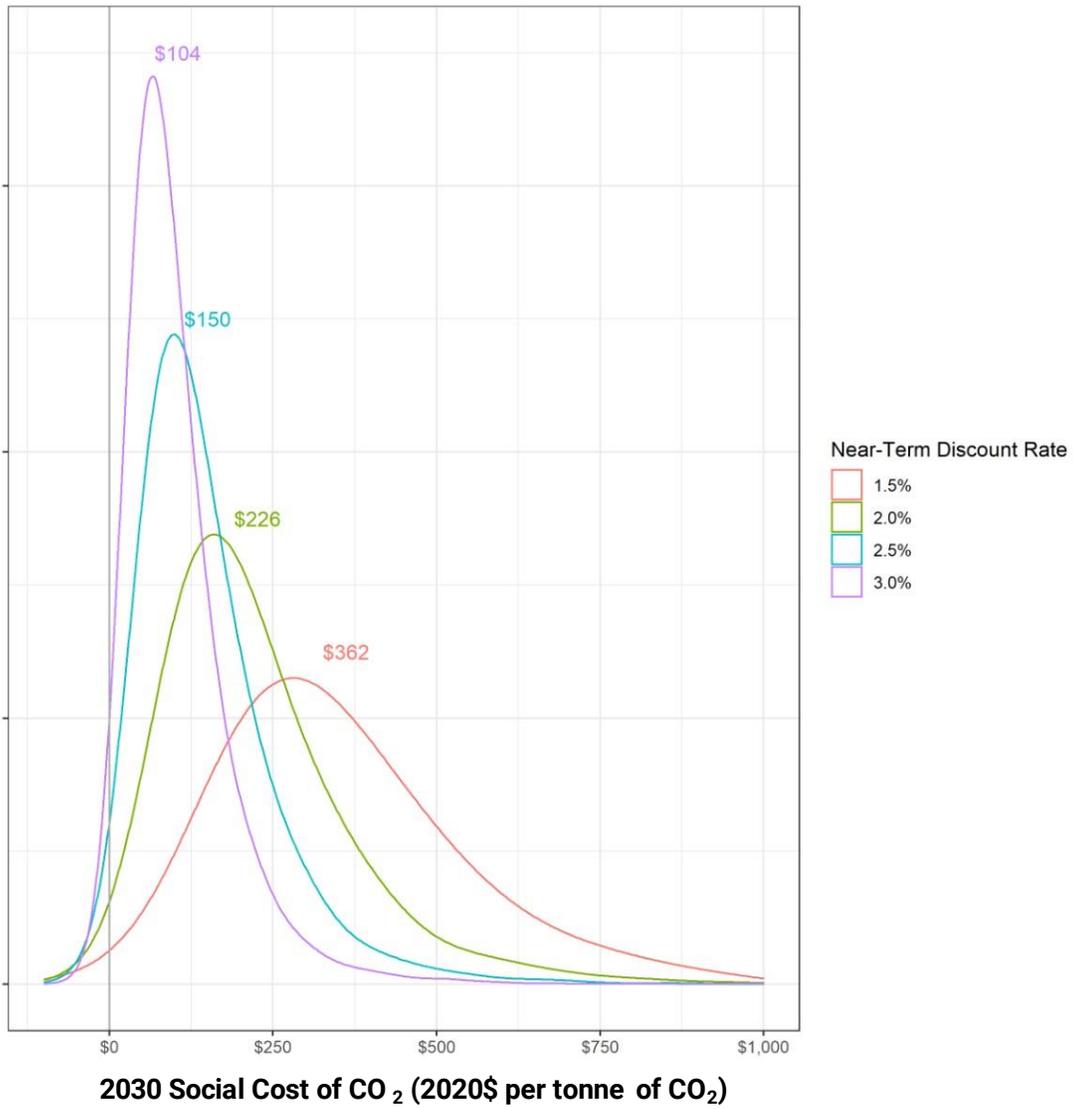

**Fig. S9. 2030 SC-CO$_2$ distributions by discount rate (2020 USD per metric tonne of CO$_2$).**
Social cost of CO$_2$ values come from Rennert, et al. (2022). Mean values are shown above each distribution with near-term average discount rates of 3.0% (purple), 2.5% (teal), 2.0% (green, preferred specification), and 1.5% (red).

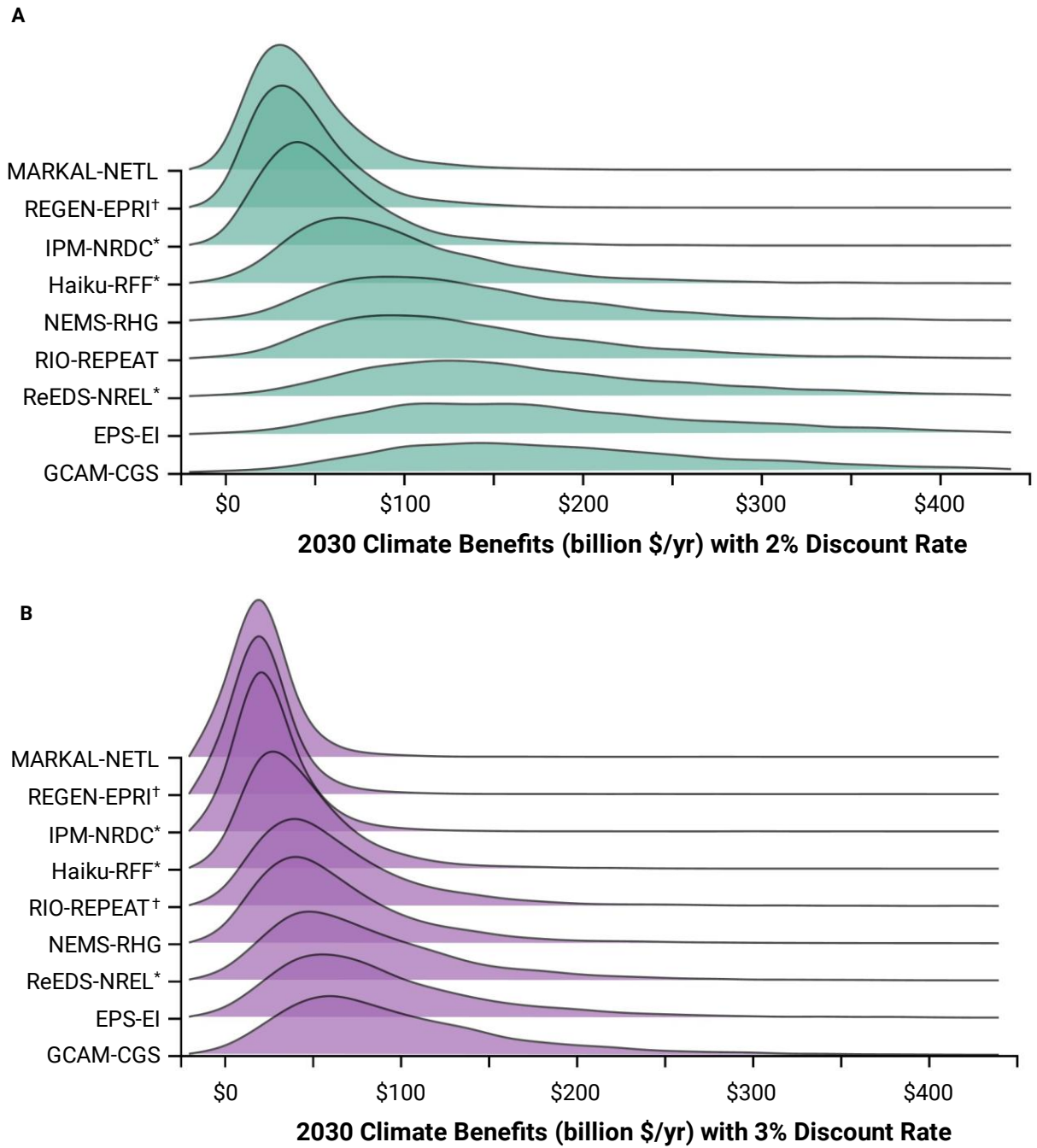

**Fig. S10. Cross-model comparison of climate benefits of IRA.** (**A**) Avoided climate-related social costs from IRA in 2030 (billion $ per year). Social cost of $CO_2$ values come from Rennert, et al. (2022) using a 2% near-term discount rate. (**B**) Climate benefits using a 3% near-term discount rate. Models with * designate that electric sector IRA provisions only are represented, and † denotes energy $CO_2$ IRA provisions only.

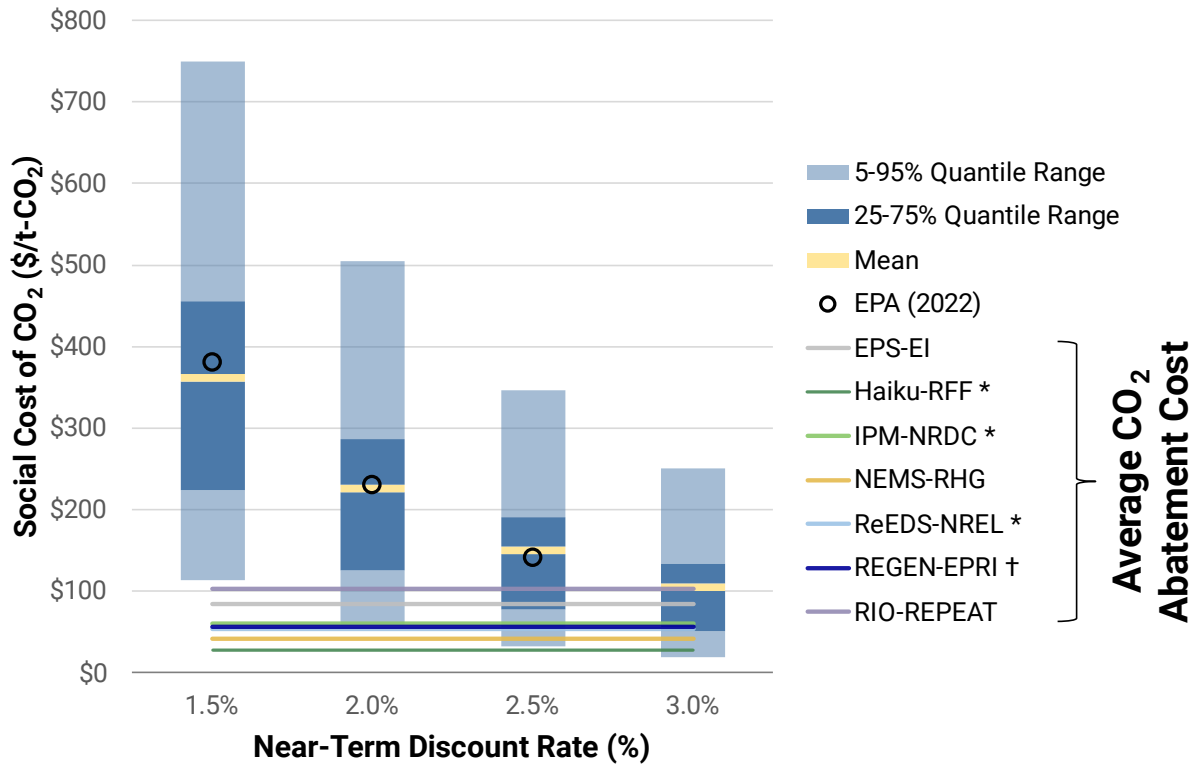

**Fig. S11. Comparison of 2030 social cost of $CO_2$ values and average abatement costs by model through 2035.** Quantile range for social cost of $CO_2$ values come from Rennert, et al. (2022). Circles show values from EPA (2022) (14). Models with * designate that electric sector IRA provisions only are represented, and † denotes energy $CO_2$ IRA provisions only.

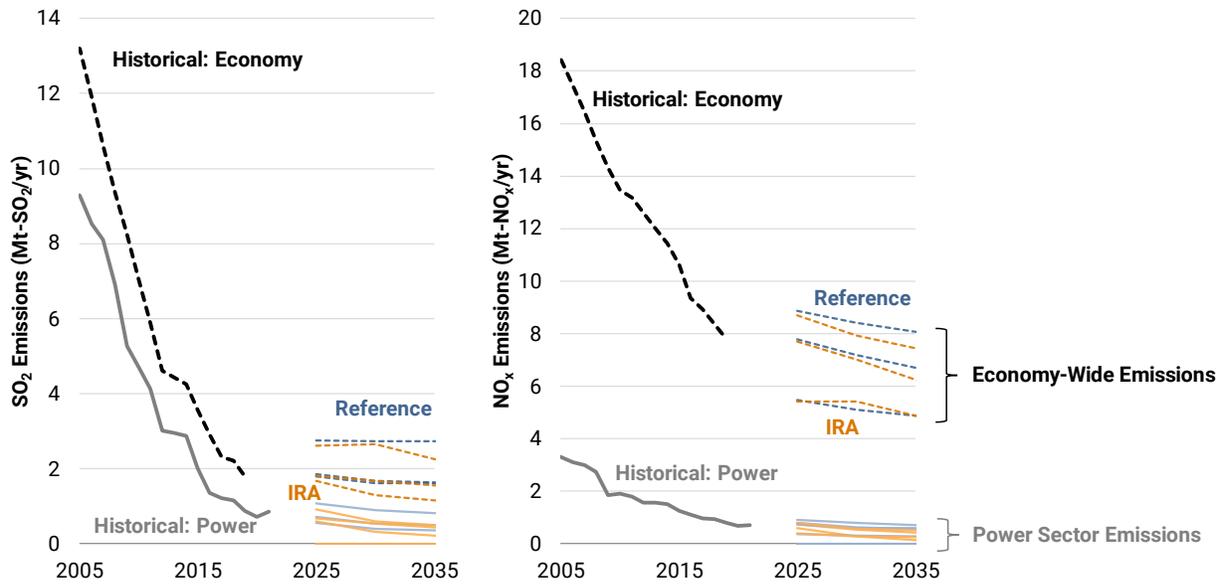

**Fig. S12. Cross-model comparison of criteria pollutant emissions by model over time.** $SO_2$ and $NO_x$ emissions (left and right panels, respectively) are shown across the economy (darker dashed lines) and for the power sector only (lighter solid lines). Historical values for economy-wide emissions come from the U.S. EPA's "Air Pollutant Emissions Trends Data" (link). Historical values for power sector emissions come from the U.S. EPA's "Power Plant Emissions Trends" (link).

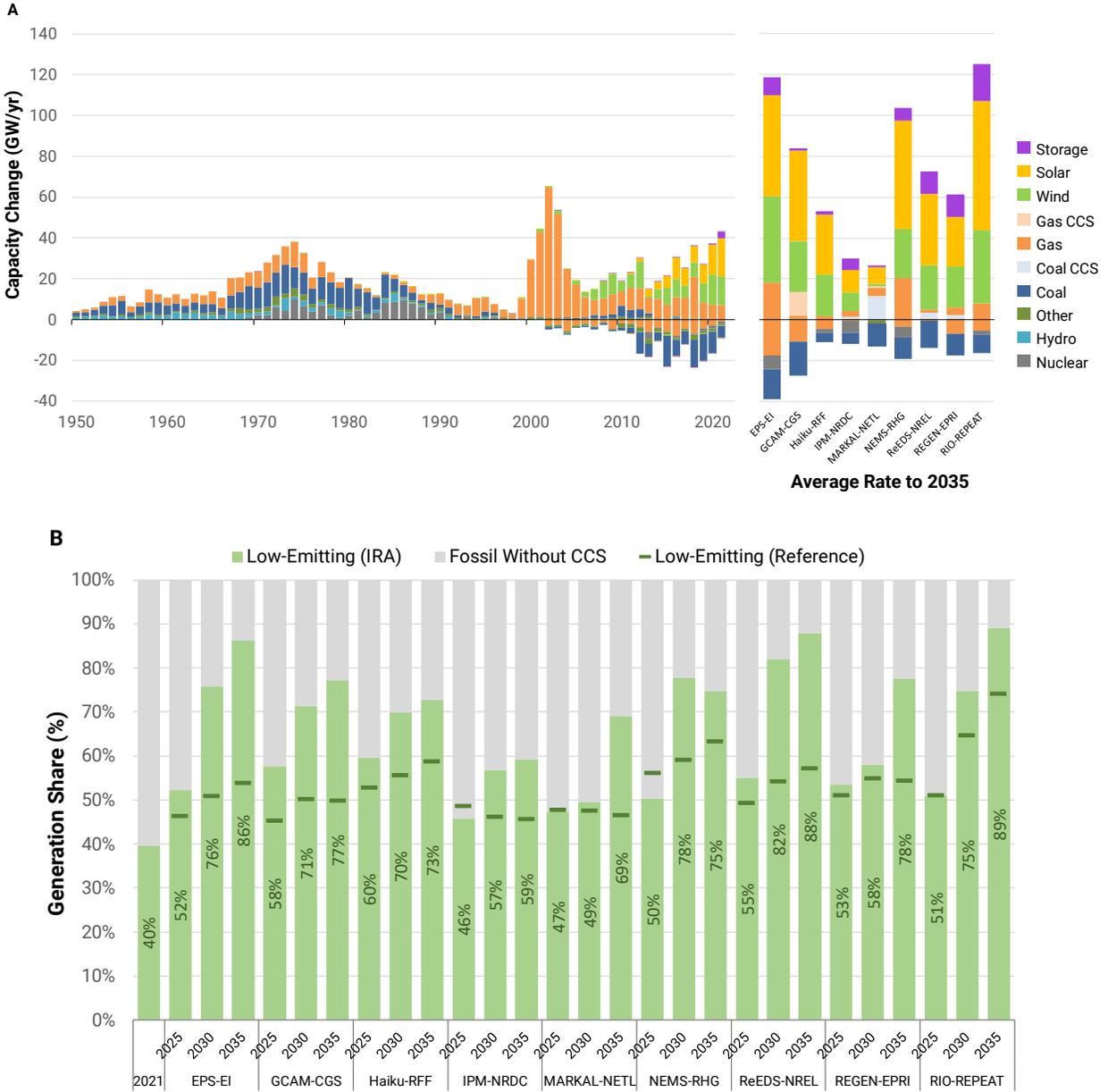

**Fig. S13. Comparison of historical and projected power sector outcomes by technology in the IRA scenarios across models.** (**A**) Additions and retirements—historical and average annual projections by model through 2035. Solar values are in $GW_{AC}$ terms and include utility-scale and distributed capacity. Historical values come from Form EIA-860 data.[17] (**B**) Low-emitting generation shares, including renewables (with biomass), nuclear, and CCS-equipped generation. Values are shown over time with IRA relative to the reference scenario.

---

[17] EIA-860 data reflect retirements since 2002. Note that additions and retirements of similar capacity can occur in the same model, which reflects differences in regional compositions (e.g., retiring capacity in some regions but adding capacity in others), technology compositions within the same fuel category (e.g., natural gas capacity includes both combined cycles and combustion turbines), as well as instances where replacing existing assets with newer capacity can lower system costs (e.g., retiring plants where going-forward costs exceed system benefits).

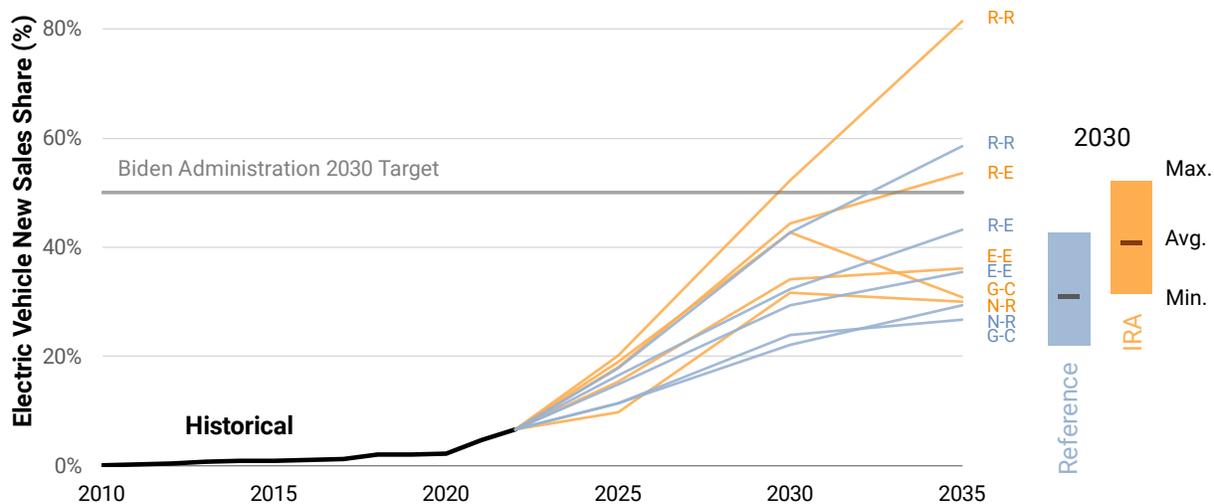

**Fig. S14. Electric vehicle new sales share of U.S. light-duty cars and trucks across models.**
Values include battery electric vehicles and plug-in hybrid electric vehicles. Historical values are from the International Energy Agency's "Global EV Outlook 2022" and Argonne National Laboratory's "Light Duty Electric Drive Vehicles Monthly Sales Update" through Dec. 2022.

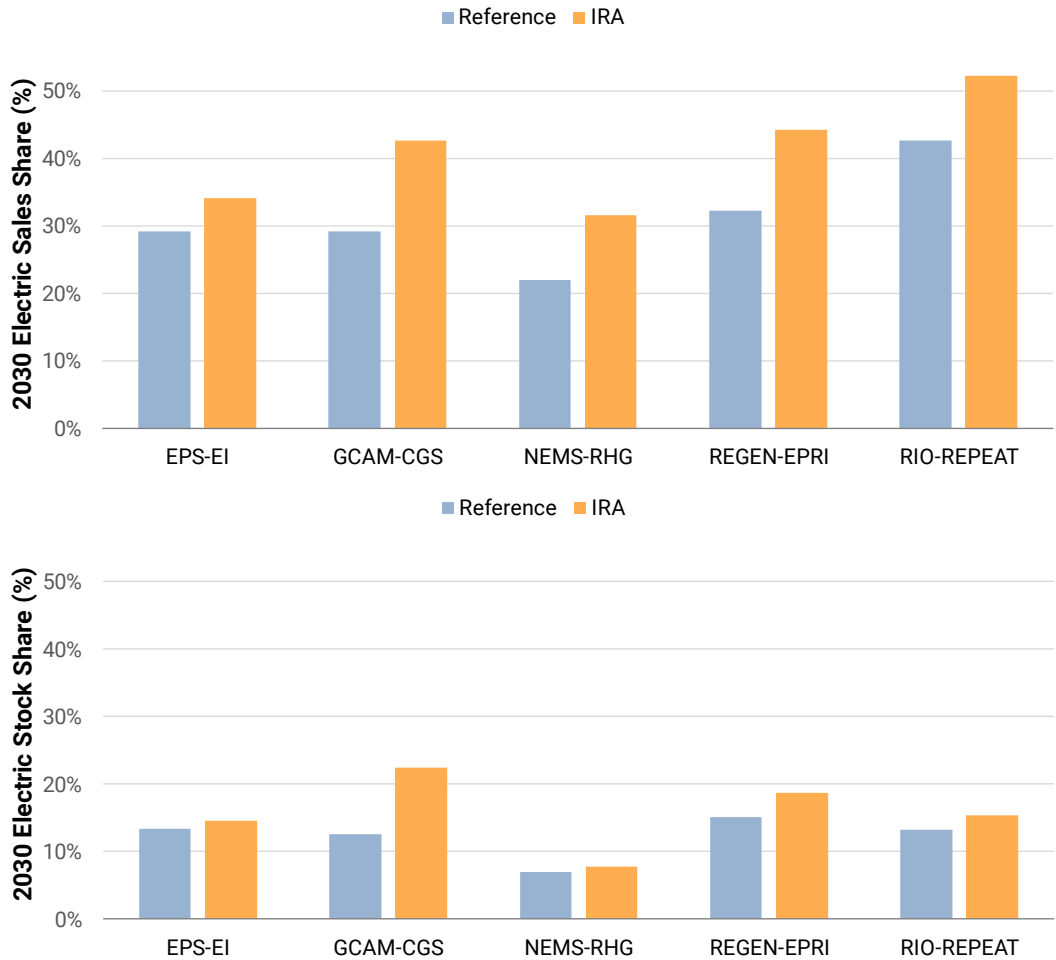

**Fig. S15. Cross-model comparison of electric vehicle shares in 2030.** New sales shares (top panel) and total vehicle stock (bottom panel) for U.S. passenger vehicles are shown for a reference scenario ("Ref") and a scenario with IRA incentives ("IRA").

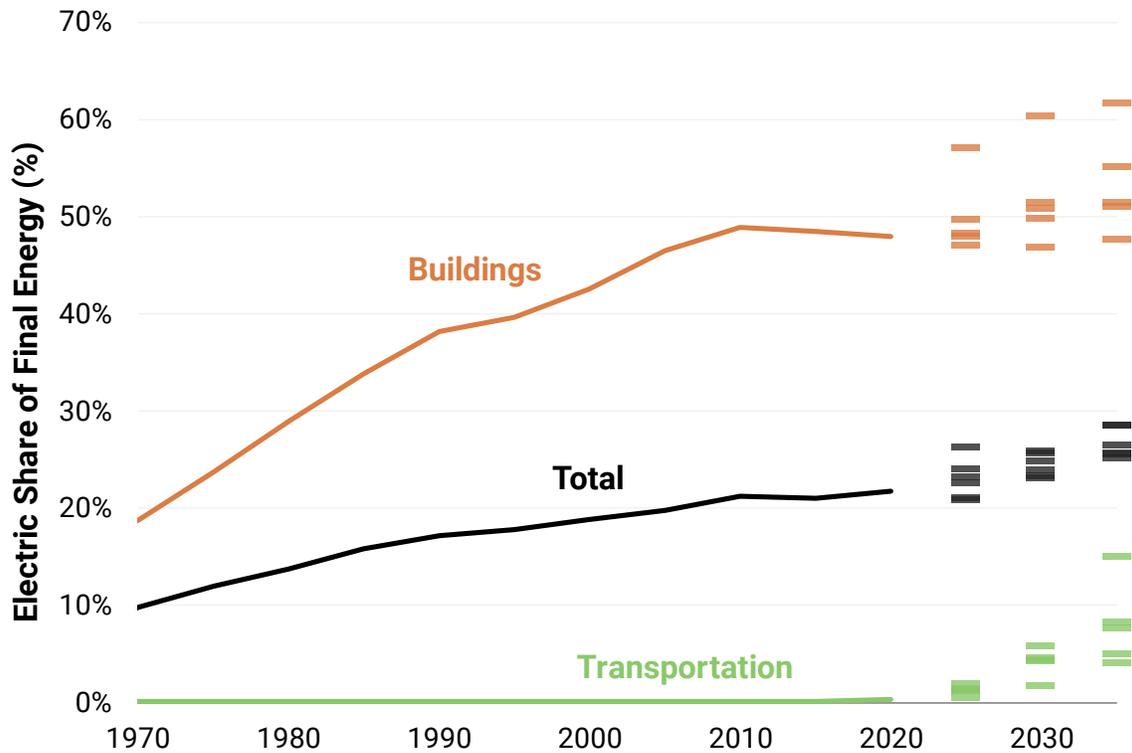

**Fig. S16. Cross-model comparison of sectoral electrification trends.** Electric share of final energy by sector under IRA scenarios shown on the right (for 2025, 2030, and 2035), where markers represent individual model results. Historical values come from U.S. EIA's "State Energy Data System" (https://www.eia.gov/state/seds/).

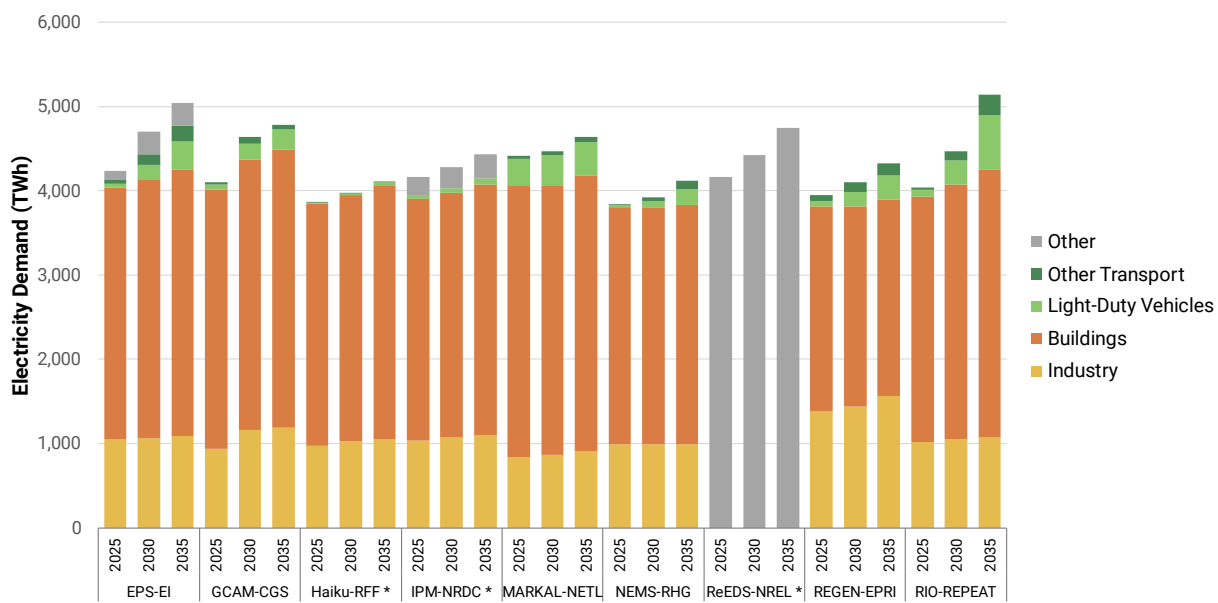

**Fig. S17. Cross-model comparison of electricity demand by sector over time.** Unspecified load is categorized as "Other." ReEDS-NREL is a power-sector-only model, so exogenous electricity demand assumptions are shown, which does not include a sectoral breakdown. Models with * designate that electric sector IRA provisions only are represented.

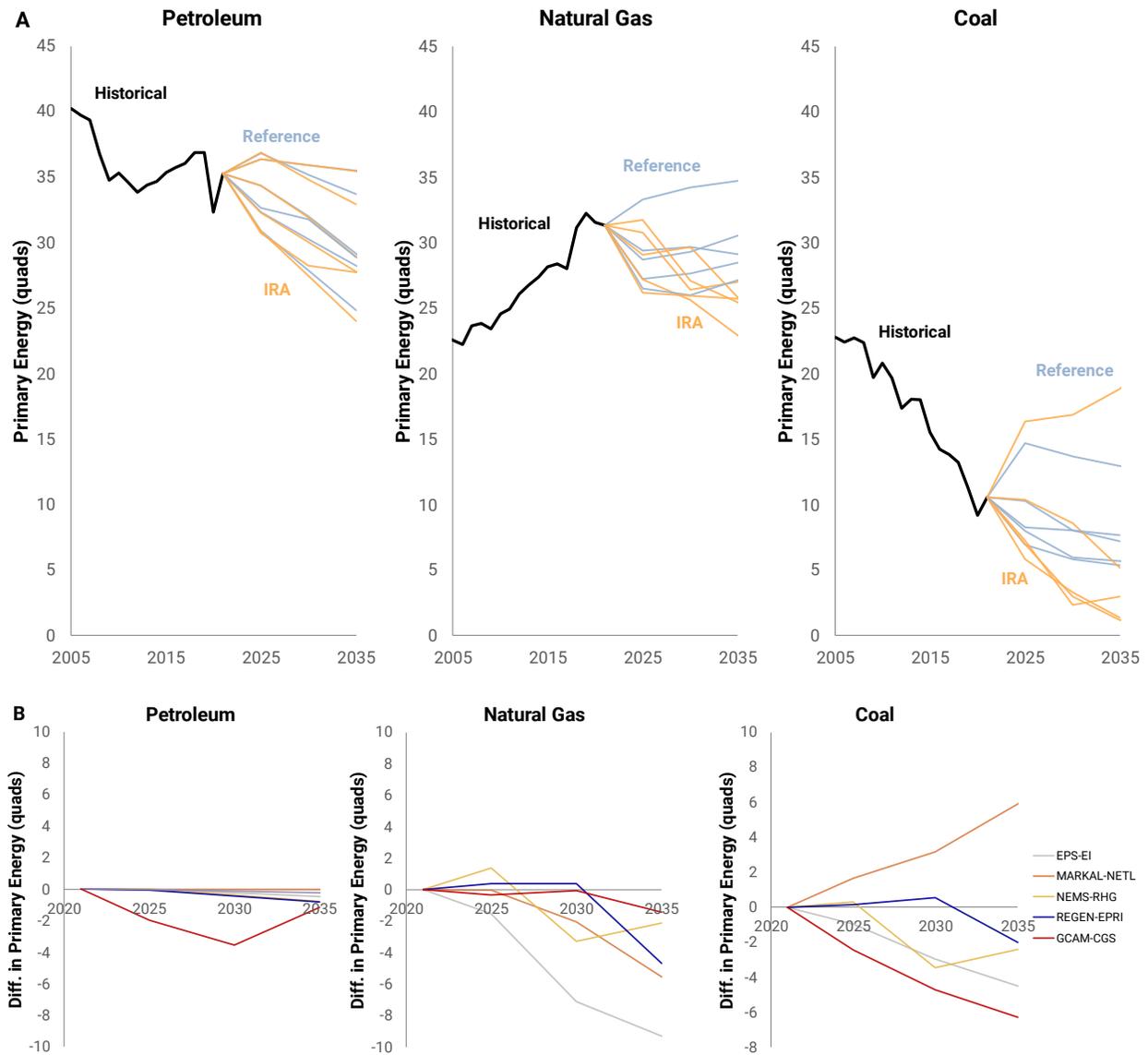

**Fig. S18. Primary energy from petroleum, natural gas, and coal across models. (A)** Values are shown for the reference scenario (gray) and IRA scenario (orange). Historical values come from the U.S. EIA's "Monthly Energy Review." **(B)** Model-specific differences in the IRA scenario relative to the reference.

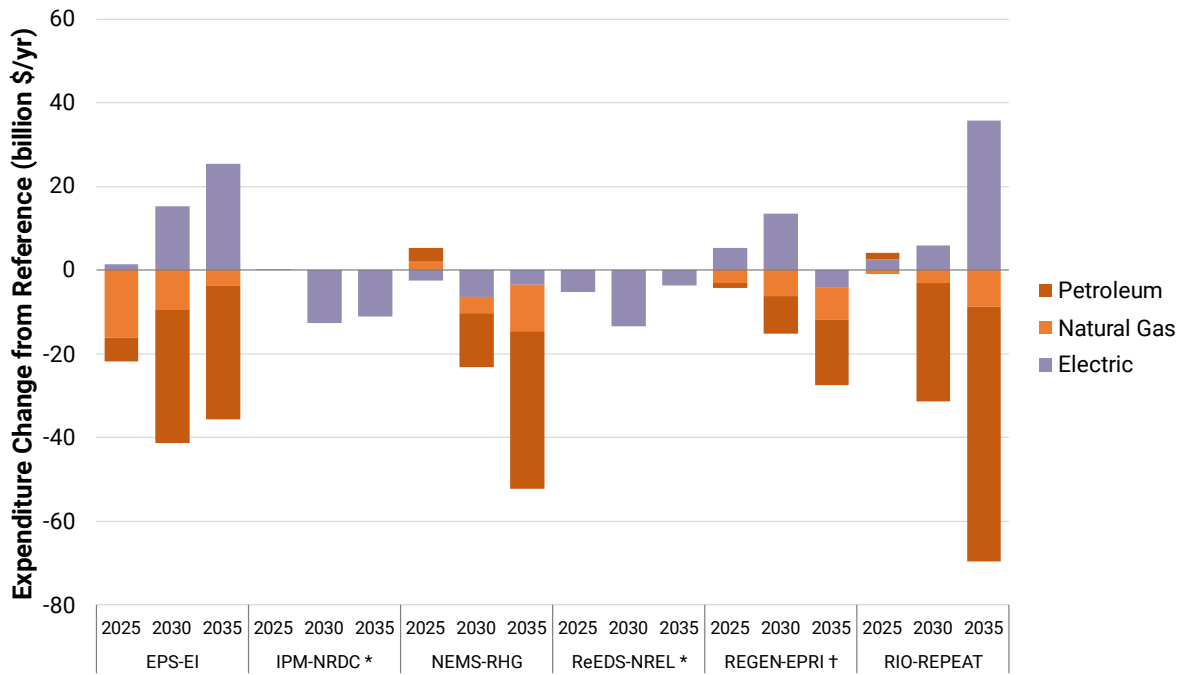

**Fig. S19. Change in economy-wide energy expenditures by fuel and model under IRA relative to reference levels.** Values shown in undiscounted real (2020) dollar terms. Models with * designate that electric sector IRA provisions only are represented, and † denotes energy $CO_2$ IRA provisions only.

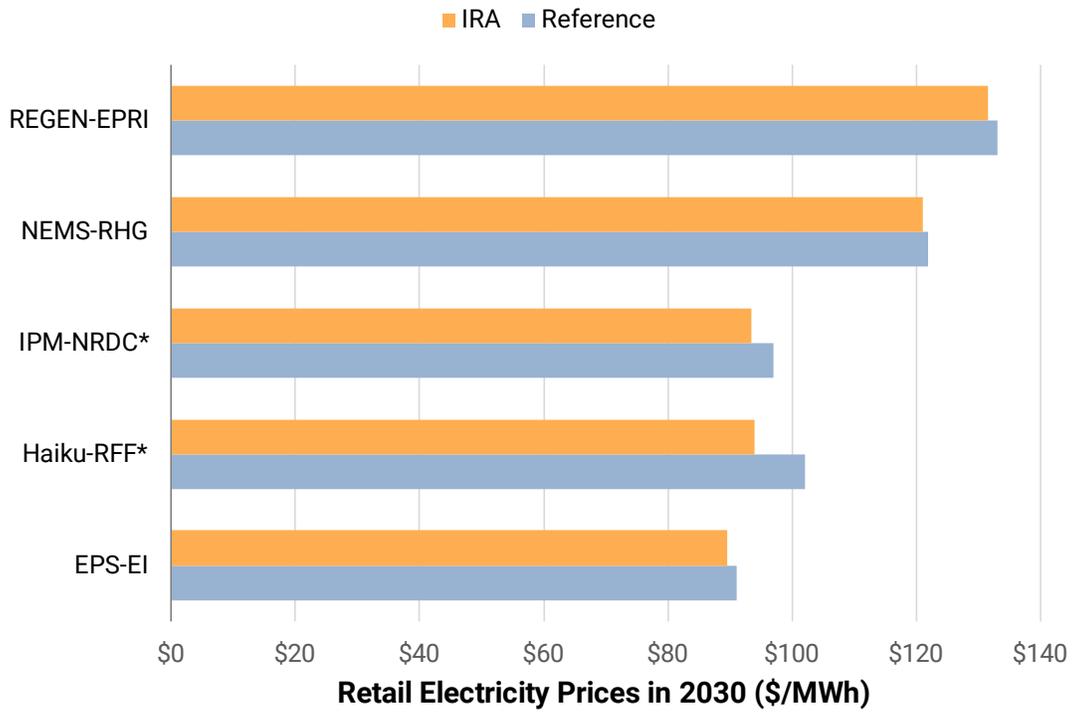

**Fig. S20. Residential retail electricity prices by model in 2030.** Values are shown for the reference scenario (gray) and IRA scenario (orange). Models with * designate that electric sector IRA provisions only are represented, and † denotes energy $CO_2$ IRA provisions only.

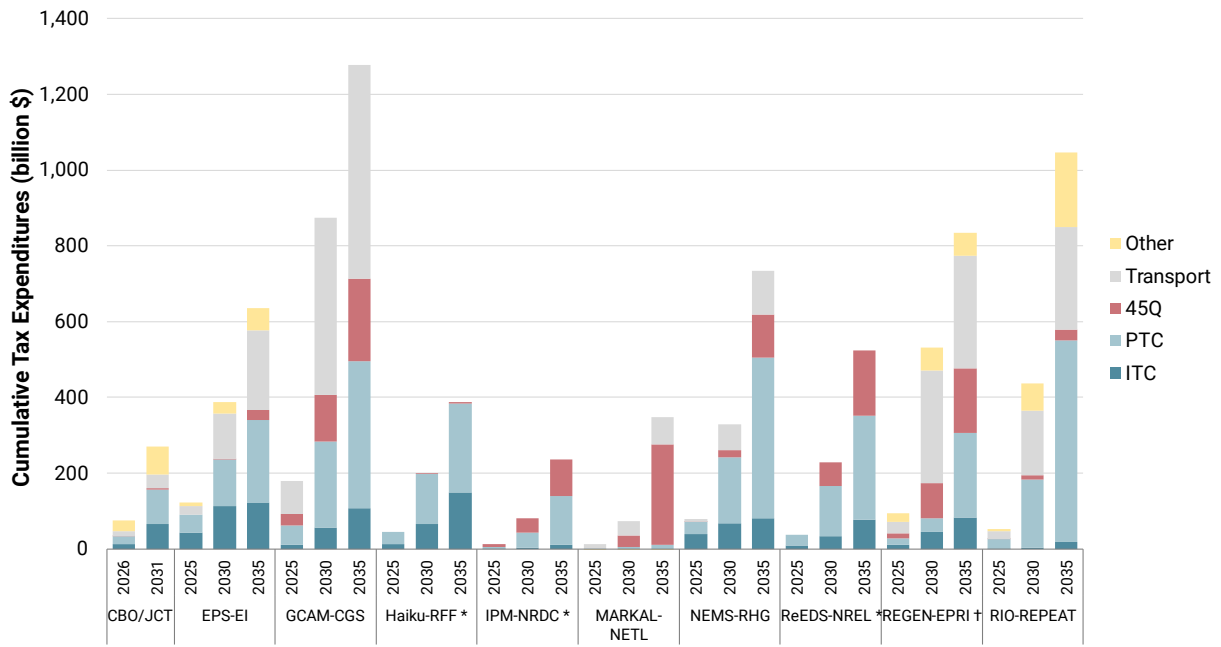

**Fig. S21. Cross-model comparison of cumulative IRA tax credit value by category over time.** Values shown in undiscounted real (2020) U.S. dollar terms. Note that Haiku-RFF, IPM-NRDC, MARKAL-NETL, and ReEDS-NREL tax credit values are shown for the power sector only. Models with * designate that electric sector IRA provisions only are represented, and † denotes energy $CO_2$ IRA provisions only. CBO/JCT estimates come from (10) and are expressed in nominal terms. Tax credits: ITC, investment tax credit (power sector); PTC, production tax credit (power sector); 45Q, credits for captured $CO_2$; 45V, credits for hydrogen.

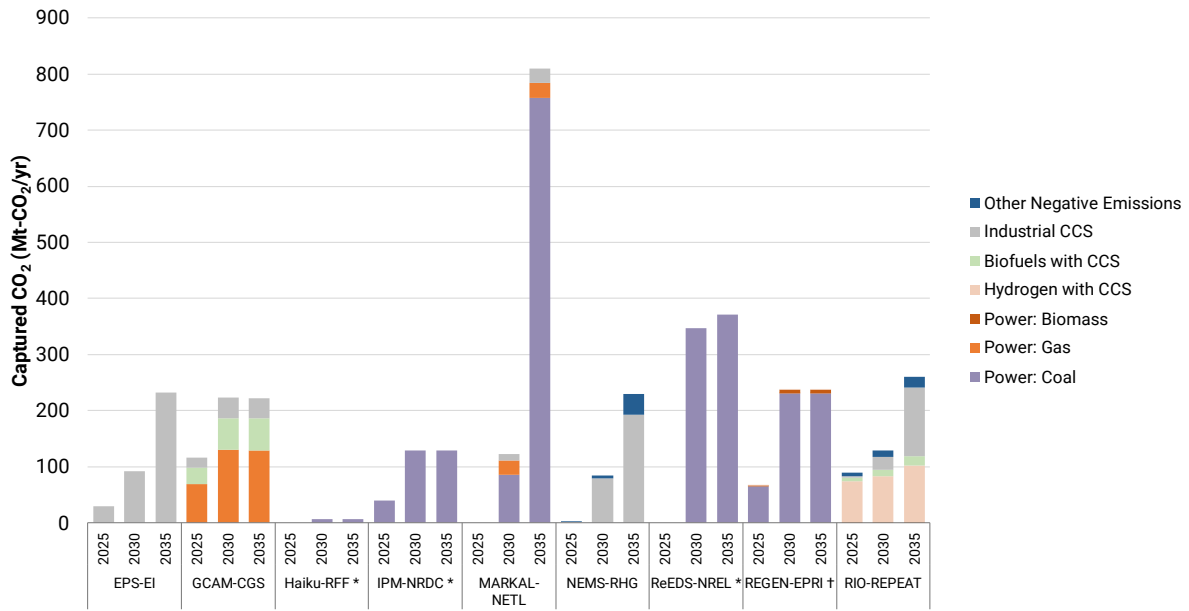

**Fig. S22. Captured $CO_2$ for transport and geologic storage under IRA.** Captured $CO_2$ is shown by category, time, and model. The "Other Negative Emissions" category is primarily direct air capture. Models with * designate that electric sector IRA provisions only are represented, and † denotes energy $CO_2$ IRA provisions only.

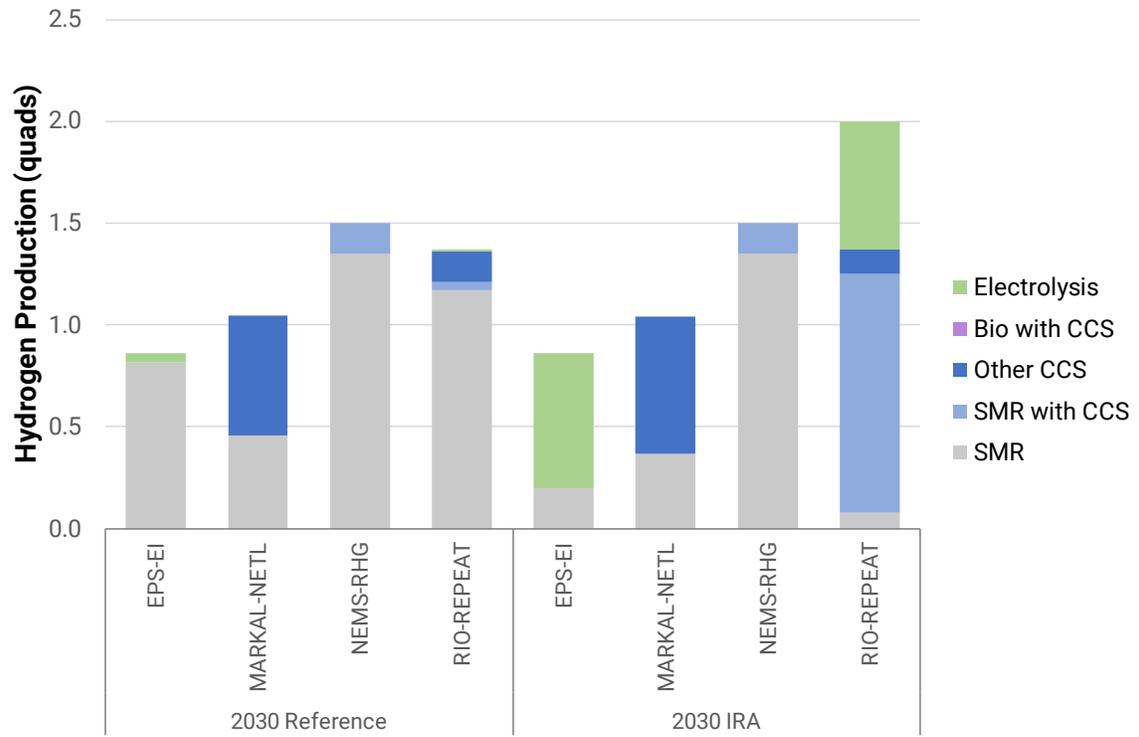

**Fig. S23. Hydrogen production by technology across models in 2030.** Values are shown for the reference scenario (left) and IRA scenario (right).

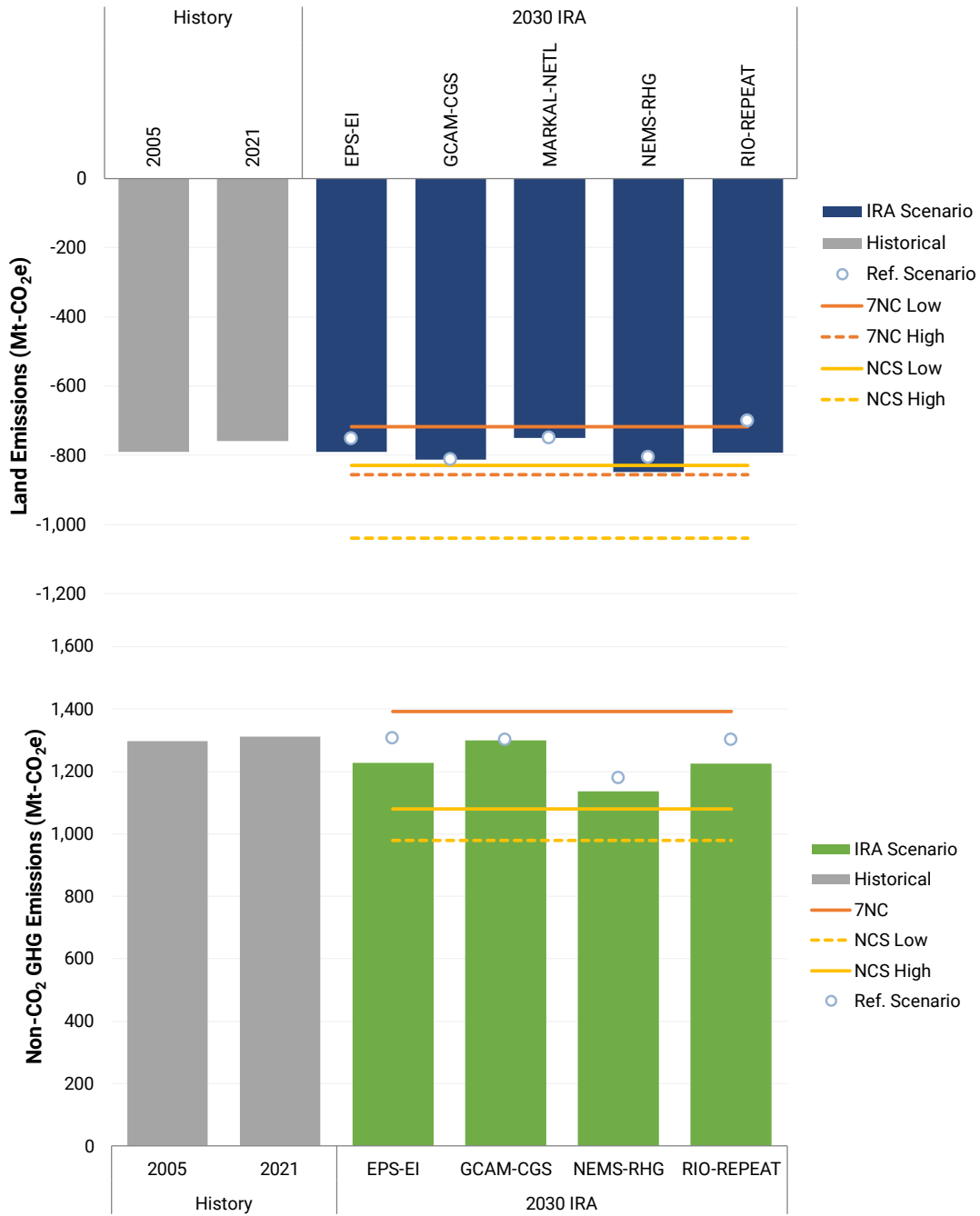

**Fig. S24. Historical and projected U.S. land sink and non-CO$_2$ GHG emissions.** Historical emissions are based on the U.S. EPA's "Inventory of U.S. Greenhouse Gas Emissions and Sinks." "7NC" lines show low and high sequestration projections from the "United States 7[th] UNFCCC National Communication, 3[rd] and 4[th] Biennial Report" reference scenarios with current policies (13). "NCS" lines show low and high sequestration projections from "The Long-Term Strategy of the United States" report's National Climate Strategy action scenarios (15).

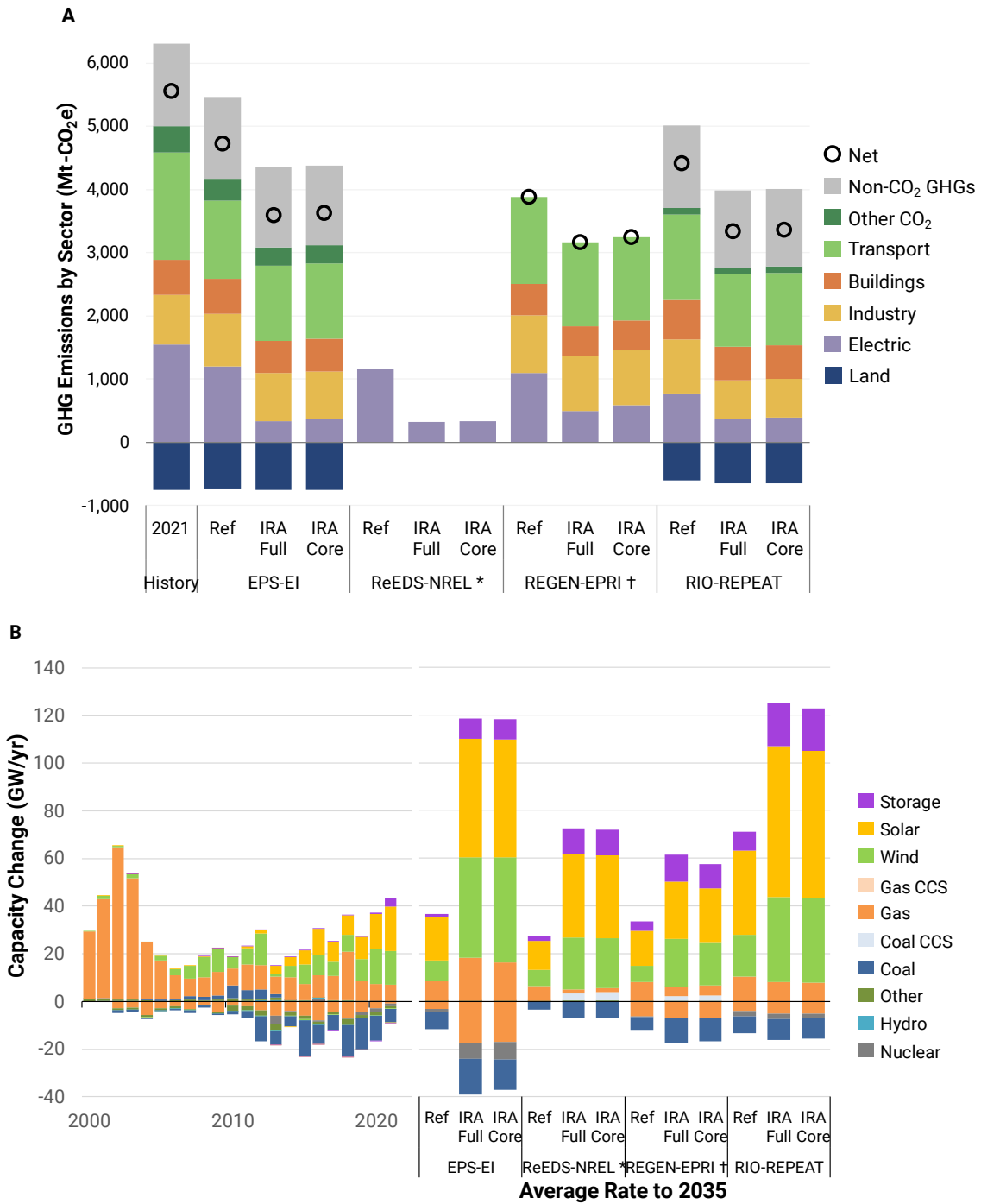

**Fig. S25. Cross-model comparison of emissions and power sector outcomes under IRA sensitivities.** (**A**) 2030 GHG emissions by sector across models and scenarios. (**B**) Electric sector capacity additions and retirements—historical and average annual projections by model through 2035. Values are shown for reference scenarios without IRA (Ref), IRA scenarios with all provisions (IRA Full), and an IRA scenario with core provisions only (IRA Core). Models with * designate that electric sector IRA provisions only are represented, and † denotes energy $CO_2$ IRA provisions only. See Section S7 for detailed scenario descriptions.

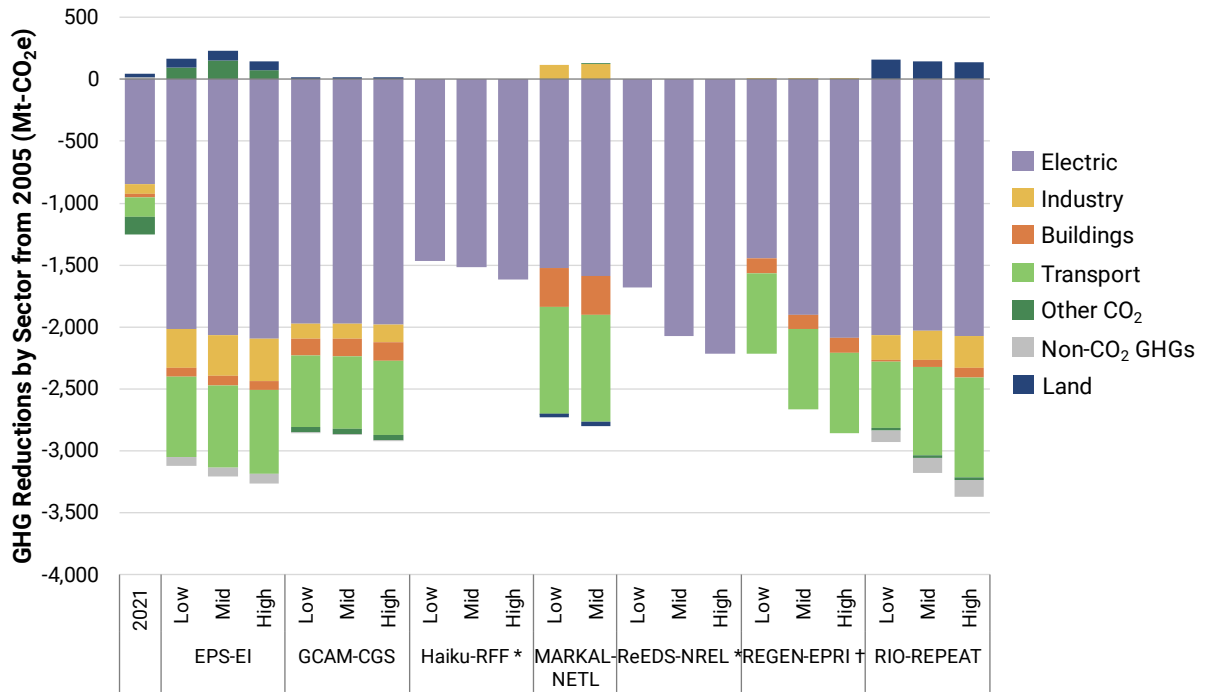

**Fig. S26. Emissions reductions by sector and model in 2035 under IRA scenarios from 2005 levels.** Low, mid, and high IRA implementation sensitivities are shown ("Mid" cases are shown in the other figures). Models with * designate that electric sector IRA provisions only are represented, and † denotes energy $CO_2$ IRA provisions only.

**Table S1.**

**Participating models and IRA provisions represented.** Section shown in parentheses. Note that the table is not an exhaustive list of IRA provisions, some of which are modeled while others are not.[18] "Not Applicable" refers to provisions that the current model structure and scope cannot represent as written.

| Sector | Program (Section) | EPS-EI | GCAM-CGS | Haiku-RFF | IPM-NRDC | MARKAL-NETL | NEMS-RHG | ReEDS-NREL | REGEN-EPRI | RIO-REPEAT |
|---|---|---|---|---|---|---|---|---|---|---|
| **Electricity** | Production tax credit (PTC) extension (13101) | Included | Included | Included | Included | Included | Included | Included | Included | Included |
| | Investment tax credit (ITC) extension (13102) | Included | Included | Included | Included | Included | Included | Included | Included | Included |
| | Solar in low-income communities (13103/13702) | Included | Not Included | | Not Included | | Included | Included | Not Included | Included |
| | PTC for existing nuclear (13015) | Included | Included | Included | Included | Included | Included | Included | Included | Included |
| | New clean electricity PTC (45Y, 13701) and ITC (48E, 13702) | Included | Included | Included | Included | Included | Included | Included | Included | Included |
| | Accelerated depreciation (13703) | | Not Included | | Included | Not Included | Included | Not Included | Included | Not Included |
| | Funds for rural coops (22004) | Included | Not Included | | | Not Included | Included | Included | Not Included | Included |
| | Transmission financing (50151) | Included | Not Included | | | | Included | Included | Not Included | Included |
| **Multi-Sector** | 45Q: Extension of credits for captured CO2 (13104) | Included | Included | | Included | Included | Included | Included | Included | Included |
| | 45V: Production credits for clean hydrogen (13204) | Included | Included | | | Included | Included | Not Included | Not Included | Included |
| | Loan authority for energy infrastructure (50144) | Included | Included | | | | Included | | Not Included | Included |
| **Transport** | Extension of incentives for biofuels (13201/13202) | Not Included | Included | | | Included | Included | | | Included |
| | Sustainable aviation credit (13203) | Included | Included | | | Included | Included | | | Included |
| | Clean vehicle credit (13401) | Included | Included | | | Included | Included | | | Included |
| | Credit for previously owned clean vehicles (13402) | Included | Included | | | | Included | | | Included |
| | Commercial clean vehicle credit (13403) | Included | Included | | | | Included | | | Included |
| | Alternative refueling property credit (13404) | Included | Included | | | | Included | | | Included |
| | Clean fuel PTC (13704) | Not Included | Included | | | | Included | | | Included |
| **Buildings** | Residential clean energy credit (13302) | Included | Included | | | Included | Included | | | Included |
| | Energy efficient commercial building deduction (13303) | Included | Included | | | Included | Included | | | Included |
| | Energy efficient home credit (13304) | Included | Included | | | Included | Included | | | Included |
| | Home energy efficiency credit (50121) | Included | Included | | | Included | Included | | | Included |
| | High efficiency home rebate program (50122) | Included | Included | | | Included | Included | | | Included |
| **Industry and Other** | Extension of advanced energy project credit (13501) | Included | Not Included | | | Included | Not Included | | | Included |
| | Advanced manufacturing production credit (13502) | Included | | | | | | | | Not Included |
| | Vehicle manufacturing loans/grants (50142/50143) | | | | | | | | | Included |
| | Advanced industrial facilities (50161) | Included | | | | | Included | | | Included |
| | Low-carbon materials (60503/60504/60506) | Included | | | | | Included | | | Included |
| | Biodiesel, Advanced Biofuels, SAF | Not Included | Included | | | Included | Included | | | Included |
| | Greenhouse Gas Reduction Fund | Included | Not Included | | | Not Included | Not Included | | | Not Included |
| | Oil and gas lease sales | Included | Included | | | | Included | | | Included |
| | Methane Emissions Reduction Program | Included | Included | | | | Included | | | Included |
| | Agriculture and forestry provisions | Included | Included | | | | Included | | | Included |

Legend: Included / Not Included / Not Applicable

---

[18] Additional energy and climate provisions include (but are not limited to) DOE ($250 billion loan authority) and USDA ($10+ billion) programs to encourage fossil-to-clean transitions (e.g., 50144, 22004); production-based manufacturing tax incentives for wind, solar, and battery equipment as well as critical materials (13502); cross-cutting programs on emissions reductions and clean technology deployment (60103, 60114); and provisions intended to support clean energy in buildings, industry, and transport (e.g., 13301, 50161, 50142, 50143).

**Table S2.**
**Participating models and key features.** Temporal resolution refers to the number of intra-annual segments.

| Analysis Abbreviation | Model(s) | Analysis Institution | Model Type | Geographic Coverage | Spatial Resolution | Temporal Resolution | Link |
|---|---|---|---|---|---|---|---|
| **EPS-EI** | Energy Policy Simulator (EPS) | Energy Innovation | Energy systems | 50 U.S. states and D.C. | Single national region | Annual for end use; seasonal for electric | Link |
| **GCAM-CGS** | Global Change Analysis Model for AP | UMD-CGS | Energy systems | 50 U.S. states and D.C. | States | Annual for end use; 4 segments for electric | Link |
| **Haiku-RFF** | Haiku Power Sector Model | Resources for the Future | Electric sector | Contiguous U.S. | States | 24 segments for electric | Link |
| **IPM-NRDC** | Integrated Planning Model | NRDC | Electric sector | Contiguous U.S. | 67 regions | 24 segments for electric | Link |
| **MARKAL-NETL** | MARKet Allocation | NETL DOE | Energy systems | Contiguous U.S. | 9 Census regions | Hourly for end use; 12 segments for electric | Link |
| **NEMS-RHG** | Rhodium Group - National Energy Modeling System | Rhodium Group | Energy systems | 50 U.S. states and D.C. | Regions vary by sector | Annual for end use; 9 segments for electric | Link |
| **ReEDS-NREL** | Regional Energy Deployment System | NREL | Electric sector | Contiguous U.S. here | 134 regions | 17 segments for electric | Link |
| **REGEN-EPRI** | Regional Economy, Greenhouse Gas, and Energy | EPRI | Energy systems | Contiguous U.S. | 16 regions | Hourly for end use; 120 segments for electric | Link |
| **RIO-REPEAT** | RIO (supply-side), EnergyPATHWAYS (demand-side) | Evolved Energy Research and ZERO Lab | Energy systems | Contiguous U.S. | 27 regions | Hourly for end use; 1,080 segments for energy supply | Link |

**Table S3.**
**Model coverage and sectoral approaches.** Coverage and equilibrium approach: PE, partial equilibrium; LP, linear program. Electric sector models are designated with * (others are energy system models, per Table S2).

| Analysis Abbreviation | Model(s) | Coverage and Equilibrium Approach | Electric Model Approach | Transport Model Approach |
|---|---|---|---|---|
| **EPS-EI** | Energy Policy Simulator (EPS) | Economy: System dynamics | Recursive dynamic[19] | Logit choice |
| **GCAM-CGS** | Global Change Analysis Model for AP | Economy: Logit choice | Recursive dynamic | Logit choice |
| **Haiku-RFF*** | Haiku Power Sector Model | Power sector PE: Least-cost LP | Perfect foresight | N/A |
| **IPM-NRDC*** | Integrated Planning Model | Power sector PE: Least-cost LP | Perfect foresight | N/A |
| **MARKAL-NETL** | MARKet Allocation | Economy: Least-cost LP | Perfect foresight | Least-cost optimization with expansion constraints |
| **NEMS-RHG** | Rhodium Group - National Energy Modeling System | Economy: 13 modules with least-cost LP supply and consumer adoption demand | Perfect foresight | Logit choice |
| **ReEDS-NREL*** | Regional Energy Deployment System | Power sector PE: Least-cost LP | User-defined (recursive used here) | N/A |
| **REGEN-EPRI** | Regional Economy, Greenhouse Gas, and Energy | Energy end use: Lagged logit choice; Electricity: Least-cost LP | Perfect foresight | Logit choice |
| **RIO-REPEAT** | RIO (supply-side), EnergyPATHWAYS (demand-side) | Economy-wide LP | Perfect foresight | Scenario-based |

---

[19] The EPS employs a simplified recursive dynamic capacity expansion model, which includes endogenously calculated changes to demand. For IRA modeling, the EPS builds on deployment estimates from ReEDS with endogenously calculated elements.

**Table S4.**
**Model representations of emerging technologies.** CCS, carbon capture and storage; $H_2$, hydrogen; T&S, transport and storage; O&M, operations and maintenance. Electric sector models are designated with * (others are energy system models, per Table S2).

| Analysis Abbreviation | CCS Technologies | $CO_2$ Transport and Storage | $H_2$ Production | $H_2$ Transport and Storage | Carbon Dioxide Removal | Energy Storage Technologies |
|---|---|---|---|---|---|---|
| **EPS-EI** | -Power: Fossil fuel (not included in IRA analysis) <br> -Industrial: Fossil fuel use and processes <br> -Direct air capture (not included in IRA analysis) | Not explicitly modeled, but costs are included in CCS costs. | $H_2$ can be produced via five different production pathways, including steam methane reforming and electrolysis. | Not modeled. | DAC: One representative technology powered by electricity | Battery storage, existing pumped hydro |
| **GCAM-CGS** | -Power: CCS for new coal, NGCC, and biomass with different capture assumptions <br> -Industrial processes <br> -Liquid fuel production | $CO_2$ T&S on a regional basis with costs for investments in pipeline and injection capacity, as well as ongoing O&M costs. | $H_2$ can be produced with electrolysis. | Exogenously specified $H_2$ transport costs. | BECCS: Power generation or liquid fuel production | Battery storage, concentrated solar power |
| **Haiku-RFF*** | Power: CCS for new coal and NGCC | EPA $CO_2$ T&S costs (step function for each state). Total $CO_2$ storage and utilization options is scaled to 100 million short tons in 2030, doubling every five years thereafter. | None | None | None | Battery storage (4-hr duration), existing pumped hydro |
| **IPM-NRDC*** | Power: CCS retrofits (90% and 99% capture) for coal and NGCC, CCS for new NGCC | Assumptions for $CO_2$ storage capacity/cost from based on GeoCAT (2021) in 37 of 48 states. $CO_2$ transport based on $228k/in-mi for pipelines. | None | None | None | Battery storage (4/8/10-hr duration), paired 4-hr battery with solar, existing pumped hydro and other storage |
| **MARKAL-NETL** | -Power: CCS for new coal, NGCC, and biomass; retrofits for coal and NGCC <br> -Industrial processes <br> -Hydrogen production <br> -Direct air capture | Fixed cost of $CO_2$ transport, injection, and long-term monitoring. $CO_2$ storage reservoir capacity varies by region. | $H_2$ can be produced with fossil resources, biomass, or electrolysis. Fossil and biomass $H_2$ technologies can be used with CCS. Local, | Transport costs from central $H_2$ vary by settlement type. Liquid $H_2$ can be imported by truck or pipeline. Distributed production technologies combine | DAC: High-temperature with heat from natural gas | Battery storage, $H_2$ storage, existing pumped hydro |

| | | | midsize, and central production options. | production and refueling capabilities. | | |
|---|---|---|---|---|---|---|
| **NEMS-RHG** | - Power: CCS for new coal and NGCC (Allam cycle); retrofits for coal and NGCC<br>-Industrial processes<br>-Hydrogen production<br>-Direct air capture | Regional $CO_2$ T&S costs | $H_2$ can be produced with fossil resources or electrolysis. Fossil can be retrofitted with CCS. | Representation of existing infrastructure. | DAC: Median cost estimate among DAC technology pathways | Battery storage, concentrated solar power, existing pumped hydro |
| **ReEDS-NREL*** | -Power: CCS for new and retrofits for coal and NGCC<br>-New biomass with CCS, DAC, and $H_2$ production modeled but not considered in this analysis | Spatially explicit cost, investment, and operation for $CO_2$ T&S, including capital and O&M of pipeline, injection, and storage. Pipelines can be built between any ReEDS regions, as well as between a region and a storage reservoir. | Available in ReEDS but not considered in this analysis. | Available in ReEDS but not considered in this analysis. | Available in ReEDS but not considered in this analysis. | Battery storage, pumped hydro storage (existing and new/uprates), compressed air, concentrated solar power |
| **REGEN-EPRI** | -Power: CCS for new coal, NGCC, and biomass with different capture assumptions; retrofits for existing coal and NGCC<br>-Industrial processes<br>-Hydrogen production<br>-Direct air capture | Regional $CO_2$ T&S with costs for investments in pipeline and injection capacity, as well as O&M costs. Investments in inter-regional $CO_2$ pipeline capacity can be made to access capacity in neighboring regions. | $H_2$ can be produced with fossil resources, biomass, or electrolysis. Fossil and biomass $H_2$ technologies can be used with CCS. | Endogenous representation of $H_2$ transport and storage with new dedicated infrastructure or blending gas commodities through existing natural gas infrastructure. | -DAC: Four representative technologies (high-temperature with heat provided by natural gas or electricity and low-temperature with gas and/or electricity)<br>-BECCS: Power generation or hydrogen production | Battery storage (endogenous duration), concentrated solar power, compressed air, $H_2$ storage, existing pumped hydro |
| **RIO-REPEAT** | -Power: CCS for new NGCC and new biomass with different capture assumptions; retrofits for existing coal and NGCC; repowering existing gas and coal to NGCC with CCS<br>-Industrial processes<br>-Hydrogen production | Inter-zonal $CO_2$ T&S through the expansion of a $CO_2$ transport network, including pipeline capital and O&M costs, injection costs, and spurline costs to connect into the trunkline system. | $H_2$ can be produced from natural gas (steam methane reformation with or without CCS, autothermal reformation with CCS), biomass with CCS or electrolysis. | Endogenous representation of $H_2$ transport with dedicated infrastructure or limited blending in existing natural gas infrastructure. Endogenous hydrogen storage technologies. | -Direct air capture<br>-BECCS: Power generation, $H_2$ production, or $H_2$ production with renewable fuel production. | Battery storage (endogenous duration), thermal energy storage, $H_2$ storage, existing pumped hydro |

**Table S5.**
**Technology-specific model expansion constraints.** See Table S4 for coverage of emerging technologies. Electric sector models are designated with * (others are energy system models, per Table S2).

| Analysis Abbreviation | Wind and Solar | Transmission | Nuclear | CCS | Other Generation Options | End-Use and Fuels |
|---|---|---|---|---|---|---|
| **EPS-EI** | None | None | None | None | None | Adoption implicitly constrained by equipment turnover |
| **GCAM-CGS** | Bounds on regional resource quality | None | Constraint on near-term deployment and state-level policies | None | None | Adoption implicitly constrained by equipment turnover |
| **Haiku-RFF*** | None | None | Fixed to baseline levels as a proxy for IRA subsidies | Upper bound on gas (coal) with CCS constrained to 20 GW (5 GW) total through 2045; upper bound on $CO_2$ storage of 100 million short tons in 2030, doubling every five years thereafter | None | N/A |
| **IPM-NRDC*** | Short-term $/kW adder for builds greater than 120% of annual record builds as of 2022 through 2035 | New transmission expansion constrained before 2028 | Economic nuclear retirements are not allowed through 2023 and are limited to 4 GW in 2025 | Only 6 GW of CCS (90% capture) can be built before 2027; 99% capture option only available starting 2027 | None | N/A |
| **MARKAL-NETL** | Bounds on regional resource quality | None | None | Upper bound on regional $CO_2$ storage reservoirs | None | Biofuel production constraints |
| **NEMS-RHG** | Bounds on regional resource quality; upper bound of 70% regional generation share | None | None | None | None | Adoption implicitly constrained by equipment turnover |
| **ReEDS-NREL*** | Bounds on regional resource quality; lower bounds for planned additions through 2024 | Near-term announced additions; before 2028, endogenous expansion is limited to historical maximum build rate (1.4 TW-mi/yr) | None | None | None | N/A |
| **REGEN-EPRI** | Bounds on regional resource quality; lower bounds for planned additions (EIA-860) | National constraint on total new transmission builds in GW-miles (10% by 2030) | Constraint on near-term deployment and state-level policies | Constraint on near-term deployment for power; bounds on $CO_2$ storage | Lower bounds to reflect under-construction capacity (EIA-860) | Adoption implicitly constrained by equipment turnover |
| **RIO-REPEAT** | Upper bound on annual builds, reflecting supply chains and interconnection, ranging from 17-30% annual growth rates through 2032 | Lower bounds on key inter-regional ties to represent the impact of DOE loans | Constrained by state-level policies | Geological sequestration: Annual limit on injection, which relaxes as the practice matures and more class six wells come online | For nascent technologies, maximum annual build constraints to reflect maturing markets, which relax over time | None |

**Table S6.**

**Summary of key indicators in 2030 for IRA and reference scenarios across models.** Indicators are 2030 electric sector $CO_2$ reductions (% from 2005 levels), 2030 generation share from low-emitting technologies (%), 2030 capacity share from low-emitting technologies (% nameplate installed capacity), 2030 coal generation decline (% from 2021 levels), economy-wide $CO_2$ reduction (% from 2005 levels), 2030 electric vehicle new sales share (% of new vehicle sold are battery or plug-in hybrid electric), 2030 electricity share of final energy (%), and 2030 petroleum reduction (% from 2005 levels). Electric sector models are designated with * (others are energy system models, per Table S2).

| Metric | Units | EPS-EI | GCAM-CGS | Haiku-RFF* | IPM-NRDC* | MARKAL-NETL | NEMS-RHG | ReEDS-NREL* | REGEN-EPRI | RIO-REPEAT |
|---|---|---|---|---|---|---|---|---|---|---|
| Ref: Electric Sector $CO_2$ Reduction | % from 2005 | 50% | 47% | 44% | 55% | 41% | 60% | 48% | 53% | 59% |
| IRA: Electric Sector $CO_2$ Reduction | % from 2005 | 75% | 72% | 63% | 66% | 47% | 80% | 83% | 58% | 69% |
| Ref: Low-Emitting Generation Share | % | 51% | 50% | 56% | 46% | 48% | 59% | 54% | 55% | 65% |
| IRA: Low-Emitting Generation Share | % | 76% | 71% | 70% | 57% | 49% | 78% | 82% | 58% | 75% |
| Ref: Low-Emitting Capacity Share | % | 50% | 48% | 51% | 47% | 52% | 53% | 49% | 53% | 58% |
| IRA: Low-Emitting Capacity Share | % | 71% | 68% | 61% | 53% | 54% | 61% | 67% | 56% | 66% |
| Ref: Unabated Coal Generation Decline | % from 2021 | 52% | 18% | 3% | 60% | 26% | 46% | 23% | 26% | 27% |
| IRA: Unabated Coal Generation Decline | % from 2021 | 79% | 76% | 37% | 78% | 43% | 85% | 92% | 48% | 47% |
| Ref: Non-Electric $CO_2$ Reduction | % from 2005 | 25% | 25% | | | 31% | 30% | | 29% | 29% |
| IRA: Non-Electric $CO_2$ Reduction | % from 2005 | 38% | 39% | | | 34% | 39% | | 33% | 37% |
| Ref: Electric Vehicle Sales Share | % | 29% | 24% | | | | 22% | | 32% | 43% |
| IRA: Electric Vehicle Sales Share | % | 34% | 43% | | | | 32% | | 44% | 52% |
| Ref: Electricity Share of Final Energy | % | 22% | 23% | | | 25% | 24% | | 24% | 23% |
| IRA: Electricity Share of Final Energy | % | 23% | 26% | | | 25% | 23% | | 26% | 24% |

| **Ref: Petroleum Decline** | % from 2005 | 25% | 21% | | | 11% | 13% | | 31% | 20% |
| --- | --- | --- | --- | --- | --- | --- | --- | --- | --- | --- |
| **IRA: Petroleum Decline** | % from 2005 | 25% | 30% | | | 11% | 14% | | 32% | 21% |